\documentclass[aps,prb,10pt,twocolumn,superscriptaddress,floatfix,longbibliography]{revtex4-2}
\usepackage[colorlinks, linkcolor=blue, citecolor=blue, urlcolor=blue]{hyperref}
\usepackage[dvipsnames]{xcolor}
\usepackage{graphicx}% Include figure files
\usepackage{dcolumn}% Align table columns on decimal point
\usepackage{amsmath}
\usepackage{mathrsfs}
\usepackage{amssymb}
\usepackage{physics}
\usepackage{braket}
\usepackage{comment}
\usepackage{textcomp}
\usepackage{upgreek}

\newcommand\numberthis[1][]{%
    \refstepcounter{equation}%
    \ifx#1\empty\else\label{eq:#1}\fi%
    \tag{\theequation}%
}
\DeclareUnicodeCharacter{00A4}{\textcurrency}
\setcitestyle{numbers,square}
\begin{document}

%\title{We don't know}
% Title brainstorming:
% * Nonlinarity? 
% * 3 wave mixing? no, too technical
% * Magnon strong coupling induced by nonlinear spin wave dynamics
% * Mediating mangon strong coupling by parametric nonlinearity
% * Magnon strong coupling induced by nonlinear parametic interactions
% * 

\title{Magnon - magnon interaction induced by nonlinear spin wave dynamics}
\author{Matteo Arfini}
\thanks{Equal contribution}
\email{m.arfini@tudelft.nl}
\affiliation{%
Kavli Institute of Nanoscience, Delft University of Technology, Lorentzweg 1,2628 CJ Delft, Netherlands
}
\author{Alvaro Bermejillo-Seco}
\thanks{Equal contribution}
\email{a.bermejilloseco@tudelft.nl}
\affiliation{%
Kavli Institute of Nanoscience, Delft University of Technology, Lorentzweg 1,2628 CJ Delft, Netherlands
}
\author{Artem Bondarenko}
\affiliation{%
Kavli Institute of Nanoscience, Delft University of Technology, Lorentzweg 1,2628 CJ Delft, Netherlands
}
\author{Clinton A. Potts}
\affiliation{Niels Bohr Institute, University of Copenhagen, Blegdamsvej 17, 2100 Copenhagen, Denmark
}
\affiliation{NNF Quantum Computing Programme, Niels Bohr Institute, University of Copenhagen, Denmark}
\author{Yaroslav M. Blanter}
\affiliation{%
Kavli Institute of Nanoscience, Delft University of Technology, Lorentzweg 1,2628 CJ Delft, Netherlands
}
\author{Herre S.J. van der Zant}
\affiliation{%
Kavli Institute of Nanoscience, Delft University of Technology, Lorentzweg 1,2628 CJ Delft, Netherlands
}
\author{Gary A. Steele}
\email{g.steele@tudelft.nl}
\affiliation{%
Kavli Institute of Nanoscience, Delft University of Technology, Lorentzweg 1,2628 CJ Delft, Netherlands
}

\date{\today}

\begin{abstract}
We experimentally and theoretically demonstrate that nonlinear spin-wave dynamics can induce an effective resonant interaction between non-resonant magnon modes in a yttrium iron garnet disk. Under strong pumping near the ferromagnetic resonance mode, we observe a spectral splitting that emerges with increasing drive amplitude. This phenomenon is well captured by a theoretical framework based on the linearization of a magnon three-wave mixing Hamiltonian, which at high power leads to parametric Suhl instabilities. The access and control of nonlinear magnon-parametric processes enables the development of experimental platforms in an unexplored parameter regime for both classical and quantum computation protocols.

\end{abstract}

\maketitle

% Introduction

Ferromagnetic resonance (FMR) \cite{griffiths1946anomalous,kittel1948theory} is a well-established experimental technique for studying the dynamics of magnons, collective spin excitations in magnetically ordered materials. The power absorption that arises when the Larmor precession of spins under an external bias field resonates with the incident microwave signal stands at the core of magnonics \cite{kruglyak2010magnonics}. Coherent interactions between spin waves and microwave fields have potential applications in sensing, transduction, and information processing \cite{Zare2022cavity,yuan2021quantum,pirro2021advances, tabuchi2015coherent, song2025single, li2020hybrid, xu2023quantum, chumak2014magnon_transistor}. The collective nature of magnons makes them well-suited for coupling to photons \cite{soykal2010strong,huebl2013high}, phonons \cite{zhang2016cavity_magnomech, potts2021dynamical}, and electronic charge \cite{bader2010spintronics}. Moreover, their integration with superconducting qubits has enabled the generation of magnon quantum states \cite{tabuchi2015coherent,lachance2017resolving,xu2023quantum}, while their tunability opens pathways to exploring non-Hermitian band theory \cite{rao2024braiding}. Theoretically, proposals exist for realizing robust magnon squeezed states \cite{kamra2016super, kamra2020magnon} and generating quantum entanglement \cite{elyasi2020resources, zhang2019quantum, pal2024using_magn_quantum_tech}.

\begin{figure}[b]
    \centering
    \includegraphics[width=\linewidth]{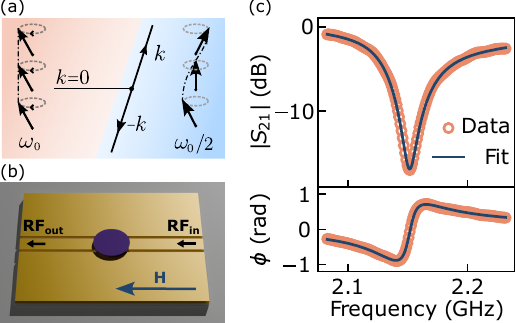}
    \caption{FMR in a driven YIG disk leading to parametric instability. (a) Schematic diagram of a magnon three-wave mixing process. A $k=0$ magnetostatic mode decays into two counter-propagating $\pm k$ modes with opposite momentum at half frequency. (b) Representation of the measured sample: YIG disk on a transmission line. (c) Measured FMR amplitude and phase response of the $k=0$ magnon mode at 28 mT. Solid lines represent a fit to the data yielding $\omega_0/2\pi=2.15$ GHz and $\gamma_{0}/2\pi=58.94$ MHz.}
    \label{fig:fig1}
\end{figure}

Most FMR experiments operate in the linear regime, where higher-order processes are negligible. However, at higher magnon amplitudes, non-linear phenomena such as Suhl instabilities can emerge \cite{suhl1957theory}. These intrinsic nonlinearities stem from higher-order contributions to the magnetic energy density in the Landau-Lifshitz-Gilbert equation \cite{pecora1988derivation} and naturally emerge from the power expansion of the spin Hamiltonian in bosonic operators under the Holstein–Primakoff transformation \cite{stancil2009spin,krivosik2010hamiltonian}. The first nonlinear term, which causes the first-order Suhl instability, involves the coupling of three magnon modes, in the form of three-wave mixing. Figure~\ref{fig:fig1}(a) schematically depicts this coupling for the case of the decay of a zero-momentum mode, $\hat{m}_0$, at frequency $\omega_0$ into two counter-propagating modes, $\hat{m}_{\pm k}$, at half the frequency $\omega_0/2$. Indirect evidence of these three-magnon interactions has been observed in studies of FMR mode saturation and decay mechanisms \cite{l2012waveturbulence,schlomann1960recent} and more recently in cavity magnonics \cite{lee2023nonlinear} and nanoscale ferromagnets \cite{barsukov2019giant,sheng2023nonlocal}. However, a systematic framework for controlling and manipulating the magnon three-wave mixing Hamiltonian has yet to be established, hindering deeper exploration and practical application of these interactions. 

In this Letter, we study a magnon-magnon coupling phenomenon originating from the three-wave interaction in a driven yttrium iron garnet (YIG) disk.  Analytical solutions of the Hamiltonian dynamics show that linearizing the intrinsic magnon three-wave mixing term at high input power yields an effective resonant beam-splitter interaction between the mode at $k=0$ and frequency $\omega_0$ and the excited magnon pair at $\pm k$ (see Figure~\ref{fig:fig1}(a)). The strength of this coupling shows a nonlinear dependence on the input microwave power. In addition, we show that, in agreement with the theoretical model, this effect is only visible below a threshold external magnetic field, above which there is no available spin wave mode at $\omega_0/2$.

% Experiment

The device consists of a 350 {\textmu}m thick, two-sided polished YIG disk with a diameter of 5 mm grown along the $[111]$ crystallographic axis by the floating zone method \cite{kimura1977single_FZ}. The disk is placed on a 50 $\mathrm{\Omega}$ transmission line narrower than the magnetic sample, as shown in Fig.~\ref{fig:fig1}(b). An external bias magnetic field $\text{H}_\text{ext}$ is applied in the plane of the disk and parallel to the feedline. We measure the microwave transmission spectrum $S_{21}$ through the transmission line, which is coupled to the disk, at room temperature using a vector network analyzer (VNA) as a function of frequency $\omega/2\pi$, input microwave power, $P$, and applied magnetic field. With a single-tone measurement, sweeping the probe in frequency as a function of external field, we observe several magnetostatic (MS) modes of the YIG disk from the transmission spectrum \cite{dillon1960magnetostatic, edwards2013magnetostatic, SI_supp}. The resonances correspond to Walker modes with azimuthal nodes along the direction of the field \cite{edwards2013magnetostatic}. Here, we address the most prominent resonance dip in the spectrum corresponding to a homogeneous mode both in-plane and across the thickness, as confirmed by micromagnetic simulations (See SI \cite{SI_supp}). The amplitude of its resulting transmission for an input power of $P=$ -20 dBm and $\text{H}_\text{ext}=$ 28 mT is shown in Fig.~\ref{fig:fig1}(c). From a fit to the FMR response of the disk \cite{S21fit}, we extract $\omega_0/2\pi=2.15$ GHz for the resonance frequency, $\gamma_{\text{ext}}/2\pi=50.06$ MHz and $\gamma_{\text{int}}/2\pi=8.88$ MHz for the external and internal mode damping rates, respectively. 

\begin{figure}
    \centering
    \includegraphics[width=\linewidth]{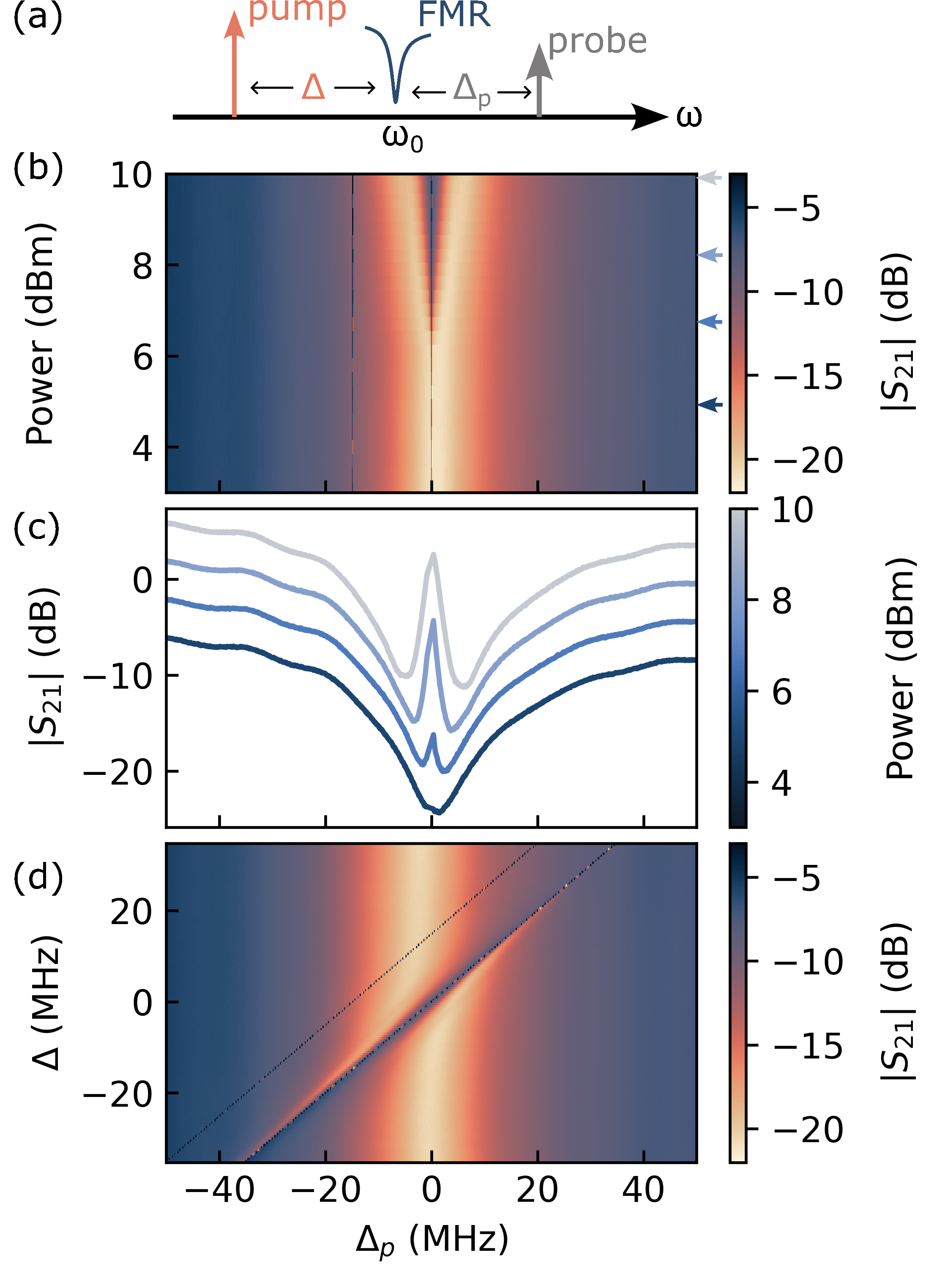}
    \caption{FMR mode splitting of a strongly driven magnon mode. (a) Two-tone measurement scheme to excite and probe 3 magnon processes. A strong pump is applied at detuning $\Delta$, and a small probe is swept with detuning $\Delta_p$. (b) Map of the transmission as a function of pump power at zero detuning ($\Delta=0$). (c) Measured transmission spectrum at different drive powers for zero detuning ($\Delta=0$). The linecuts are extracted from (b) at the powers indicated with arrows, shifted by 4dB for clarity, and the data points corresponding to the strong pump and its leak image have been filtered. (d) Map of the transmission as a function of pump and probe detunings for power 10 dBm.}
    \label{fig:fig2}
\end{figure}

To study nonlinear parametric processes, we adopt a two-tone measurement scheme, as depicted in Fig.~\ref{fig:fig2}(a). A strong pump tone is applied with a small detuning $\Delta$ from the main $k=0$ mode resonance, while a weak probe with detuning $\Delta_p$ is swept across the FMR mode spectrum to measure its response. First, we focus on the case where the pump detuning $\Delta$ is zero. As the pump power is increased above a critical value, the $k=0$ FMR response splits into two separate resonances with equal amplitude (see Fig.~\ref{fig:fig2}(b-c)). When the pump power is fixed at a value greater than the threshold and the pump detuning $\Delta$ is swept across the $k=0$ mode, the resulting spectrum shows the signature of normal mode splitting (see Fig.~\ref{fig:fig2}(d)). 
% Discussion

The experimental data can be accurately represented by the microscopic theory of nonlinear spin-wave dynamics \cite{stancil2009spin,l2012waveturbulence,suhl1988spin}, achieved through the linearization of the magnon three-wave mixing process under a strong pump drive. We model our system starting from the Hamiltonian in frequency domain $\mathcal{\hat{H}}=\mathcal{\hat{H}}_0+\mathcal{\hat{H}}_{\text{int}}+\mathcal{\hat{H}}_d$, in which
\begin{equation}
    \mathcal{\hat{H}}_0/\hbar=\omega_0\hat{m}^\dagger_0\hat{m}_0+\sum_{k>0}\bigg(\omega_k\hat{m}^\dagger_k\hat{m}_k + \omega_{-k}\hat{m}^\dagger_{-k}\hat{m}_{-k}\bigg)
\end{equation}
describes the bare resonant terms of both the $k=0$ magnon mode probed by FMR, and $k\neq 0$ spin wave modes in the disk, with $\omega_k=\omega_{-k}$. The RF contribution at $\omega_d$ from the input microwave power to the feedline is enclosed in the term $\hat{\mathcal{H}}_d/\hbar=i(\Omega_d^*\hat{m}^\dagger_0 e^{i\omega_d t}-\Omega_d\hat{m}_0 e^{-i\omega_d t})$, where $\Omega_d$ represents the coherent amplitude (in Hz) of the driving field. The three-magnon scattering is captured by the nonlinear interaction term as derived by Suhl \cite{suhl1957theory},
\begin{equation}
    \mathcal{\hat{H}}_{\text{int}}/\hbar=\sum_{k>0}V_k\hat{m}^\dagger_0 \hat{m}_k\hat{m}_{-k} + h.c.,
\end{equation}
where the coupling $V_k$ depends on sample parameters, and scales inversely proportional to the number of spins \cite{stancil2009spin,l2012waveturbulence}. Higher-order nonlinearities are not considered as they are not necessary to explain the presented phenomenon. To find the effective interaction between the ($k$,$-k$) magnon pair and the $k=0$ mode, we further analyze the presented dynamics. 

\begin{figure}[t]
    \centering
    \includegraphics[width=0.9\linewidth]{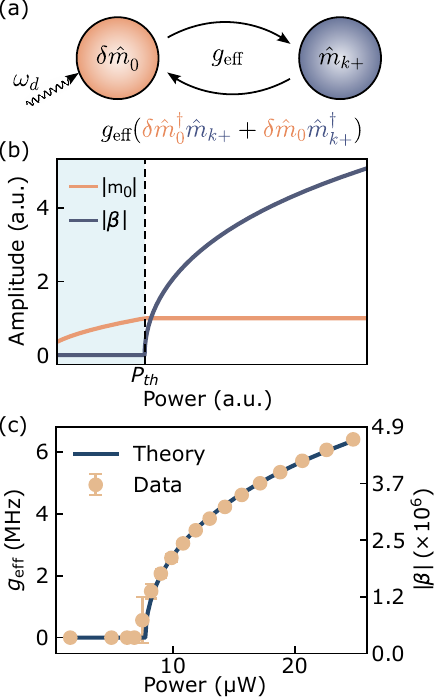}
    \caption{Effective power dependent beam splitter interaction between the driven $k=0$ mode and the parametrically excited magnon pair. (a) Interaction scheme between the fluctuations of the strongly driven $k=0$ mode and the $\pm k$ mode. (b) Power dependence of the magnon mode amplitudes. The dashed line indicates the saturation power $P_{th}$. The amplitude values are normalized to the saturation amplitude of $\hat{m}_0$. (c) Power dependence of the extracted coupling $g_{\text{eff}}$  from measurement data at 30 mT, fitted with Eq. (\ref{eq:beta_amplitude}). The right scale indicates the correspondence with the steady state amplitude of mode $\beta$.}
    \label{fig:fig3}
\end{figure}

\begin{figure*}[t]
    \centering
    \includegraphics[width=0.9\linewidth]{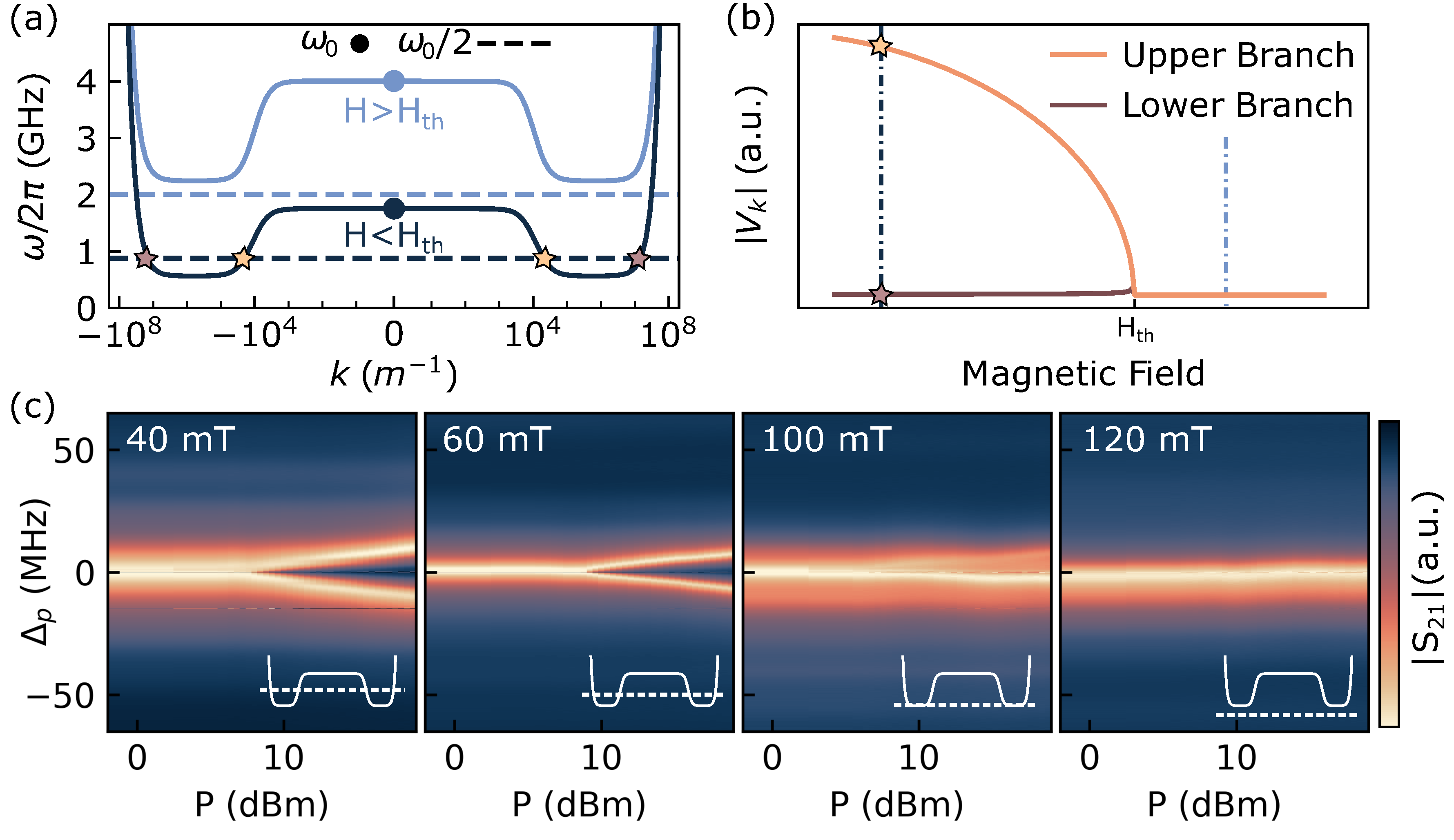}
    \caption{Magnetic field dependence of the splitting. (a) Calculated dispersion relation for two magnetic fields below and above the threshold. Below the threshold, there are two pairs of available states at $\omega_0/2$ (dashed line) indicated with stars. (b) Calculated magnetic field dependence of the coupling $V_k$. The upper (lower) branch corresponds to the available state of lower (higher) $k$. Details of the calculation can be found in \cite{SI_supp}. (c) Power dependence at zero pump detuning ($\Delta=0$), for magnetic fields of 40, 60, 100, and 120 mT. In the lower right corner, a schematic illustrates the dispersion relation and the corresponding position of $\omega_0/2$.}
    \label{fig:fig4}
\end{figure*}

% Power dependence

After removing the time dependence from the Hamiltonian by going into a frame rotating at frequency $\omega_d$, we derive the Heisenberg equations of motion for the modes and solve for the corresponding steady-state amplitudes. Under mean-field approximation, and considering $\vert\langle \hat{m}_{k} \rangle\vert = \vert\langle \hat{m}_{-k} \rangle\vert =  \beta $, the amplitude of the coherently driven $k=0$ mode (at $\Delta=0$) reaches a threshold for increasing $\Omega_d$ (see the SI \cite{SI_supp}),
\begin{equation}
       \langle \hat{m}_0\rangle_{cr} = \frac{\Omega_d}{\gamma_{0}/2} = \frac{\gamma_k}{2V_k},
\end{equation}
with total loss rates, $\gamma_0$, and $\gamma_k$ of the $k=0$ and downconverted modes, respectively. Since the final expression for $\langle \hat{m}_0\rangle_{cr}$ is power independent, the number of excited magnons at $k=0$ cannot exceed the critical value $(\gamma_k/2V_k)^2$, which is solely determined by the system parameters. Together with a saturation of the $k=0$ mode, the $k\neq0$ mode experiences the Suhl parametric instability and acquires a coherent amplitude,  which for zero detuning is given by
\begin{equation}
    \beta = \sqrt{\frac{4V_k\Omega_d-\gamma_{0}\gamma_k}{4V_k^2}}.
    \label{eq:beta_amplitude}
\end{equation}
Equation (\ref{eq:beta_amplitude}) sets a lower bound on the input power strength $\Omega_{d,cr} = \gamma_k\gamma_0/4V_k$ that is required for the $\omega_0/2$ mode to acquire a nonzero amplitude, and an upper threshold on the saturation of the main resonant mode. This condition, which is a common trait of parametric nonlinear phenomena \cite{hill1978cw_3waveoptics,zorin2016josephson}, is due to the coherent backreaction of the magnon pair on the $k=0$ once $\Omega_{d,cr}$ is crossed.

The splitting feature under a strong pump emerges by linearization of the equations of motion for small fluctuations $\delta m_{0,\pm k}$ around the steady-state solution (see SI \cite{SI_supp}), via the same three-wave mixing term that led to the Suhl instability. The resulting interaction term in $\mathcal{\hat{H}}$ can be recast in a simple form by performing a Bogoliubov transformation on the $\pm k\neq 0$ modes. We find that
\begin{equation}
    \mathcal{\hat{H}}'_{\text{int}}/\hbar=\sqrt{2}V_k\beta(\delta \hat{m}_0 \hat{m}_{k+}^\dagger+\delta \hat{m}^\dagger_0 \hat{m}_{k+})+\mathcal{\hat{H}}_{SQ},
    \label{beam_splitter_H}
\end{equation}
where in the new canonical basis we have introduced the mode $\hat{m}_{k+} = (i\delta\hat{m}_k + \delta\hat{m}_{-k})/\sqrt{2}$. Equation (\ref{beam_splitter_H}) describes a resonant beam splitter interaction between the $k=0$ mode and the pair of parametrically excited $\pm k$ magnon modes. The effective coupling $g_{\text{eff}}=\sqrt{2}V_k\beta$ scales proportionally with the coherent amplitude of the excited magnon pair, and exhibits a nonlinear dependence on the input RF pump power. The second term $\mathcal{\hat{H}}_{SQ}$ contains a single-mode squeezing contribution for mode $\hat{m}_{k+}$ and its orthogonal mode, and is a direct consequence of the magnon three-wave mixing process. The orthogonal mode is not coupled to $\hat{m}_0$ as shown in the SI \cite{SI_supp}.

Figure~\ref{fig:fig3}(b-c) highlights the consistency between the theoretical model and experimental findings. As the pump power increases, the population of the driven mode rises until it reaches a threshold, as shown in Fig.~\ref{fig:fig3}(b). At this point, the amplitude of $\hat{m}_0$ saturates, and downconverted modes with $\pm k$ and amplitude $|\beta|$ are coherently excited by the strong drive. This phenomenon is evident in the experimental data, where the observed spectral splitting is proportional to $|\beta|$.

To demonstrate this proportionality we extract the empirical effective coupling $g_{\text{eff}}$ from the splitting feature in the $S_{21}$ two-tone spectrum (see SI \cite{SI_supp}) and then fit the results to Equation (\ref{eq:beta_amplitude}). Here, the driving rate is related to the input microwave power, using standard input-output theory: $\Omega_{d,cr}=\sqrt{P_{th}\gamma_{\text{ext}}/2\hbar\omega_0}$. The only free parameters in the fit are the attenuation of the input RF signal in the measurement setup and the loss term $\gamma_k$. From the fitting, we obtain a value of $\gamma_k=1.77$ MHz for the $\pm k$ magnon pair mode and a total attenuation of 28 dB due to the coaxial cables and electronic instruments. Using the expression for circularly polarized waves \cite{l2012waveturbulence}, we compute the bare three-magnon scattering strength, $V_k$, and find a value of 0.91 Hz. These calculations, illustrated in Fig.~\ref{fig:fig4} (a-b) and detailed in the SI \cite{SI_supp}, align well with the retrieved $g_{\text{eff}}$, confirming the theoretical power dependence of the downconverted mode amplitude, $|\beta|$. This agreement establishes that the resonant three-magnon interaction emerges only for input powers exceeding the threshold. Similar experimental evidence was observed in a recent study conducted on a YIG sphere driven at high power \cite{rao2023unveiling}. The arising of the splitting was interpreted via a phenomenological model involving an auxiliary dark magnon mode resonant with the pump.

% Magnetic field dependence
To further support our interpretations of the experimental observations, we study the magnetic field dependence of the measured spectral splitting. The downconversion of $\hat{m}_0$ to $\hat{m}_{k+}$ relies on the availability of states at energy $\omega_0/2$ with momenta $\pm k$. An external magnetic field modifies the magnon dispersion curve such that the availability of states becomes field-dependent. We theoretically investigate under which external field this requirement can be met. For that, we approximate the dispersion relation of the in-plane magnetized disk by that of an ellipsoid, which has a uniform demagnetization field (see details in SI \cite{SI_supp}). The dispersion relation of spin waves propagating along the applied field is shown in Fig. \ref{fig:fig4}(a), for two different values of the external field, above and below a threshold field, $H_{th}$. By indicating the value of $\omega_0/2$ with a dashed line, we show that for $H<H_{\text{th}}$ there are two pairs of available states marked with stars, while for $H>H_{\text{th}}$ there are none. In the SI \cite{SI_supp}, we make use of the formalism for computing $V_k$ developed by \cite{l2012waveturbulence}, also for a uniform demagnetisation field, and arrive at two conclusions: 1) in the case where there are two possible pairs, the coupling of the lower $k$ pair is 3 orders of magnitude higher, enabling us to disregard the higher $k$ pair, and 2) as the external field is increased the rate diminishes until it vanishes at the point where the energy momentum conservation condition can no longer be met. These two conclusions are illustrated in Fig.~\ref{fig:fig4}(b), where we show $V_k$ for the two branches with lower and higher $k$ (upper and lower branches, respectively). The coupling in the lower branch is negligible compared to the upper one up to the threshold. 

The power dependence at zero pump detuning ($\Delta=0$) for four different magnetic fields is presented in Fig.~\ref{fig:fig4}(c). At 40 and 60 mT, the splitting is clear, with a slightly higher power threshold at 60 mT. At 100 mT, the splitting is barely visible, and at 120 mT, there is no splitting at all. These results yield an experimental threshold field, $H_{\text{th}}$, between 100 and 120 mT. Theoretically, as detailed in the SI \cite{SI_supp}, an ellipsoidal sample approximation predicts $H_{th}=58$ mT, approximately a factor of 2 lower than the experimental value. We attribute this discrepancy to simplified assumptions in calculating the dispersion relation and $V_k$, particularly the neglect of inhomogeneous demagnetization fields arising from the sample's cylindrical geometry, which affects dipolar dispersion. Notably, the demonstrated magnetic field dependence of $V_k$ provides a direct experimental handle for tuning the strength of the beam-splitter interaction, enabling precise control of magnon dynamics.

%Conclusion

In summary, we have observed the strong interaction of a magnetostatic mode with a dark mode under strong pumping, revealing a novel scheme to achieve magnon-magnon interactions that could advance our understanding of nonlinear wave-mixing phenomena in spin systems. Starting from the spin bosonic Hamiltonian model including magnon three-wave mixing, we demonstrate that the mode splitting corresponds to an excitation of a pair of magnons with opposite momentum and half frequency, leading to first-order Suhl instabilities. The model enables the extraction of the effective coupling, which is a direct measure of the power-dependent population of the magnon pair. It also predicts a threshold magnetic field for non-zero coupling, in good agreement with experimental observations.
Our results shed new light on the understanding of nonlinear parametric magnon phenomena and their dependence on bias parameters, by establishing an excellent agreement between theoretical predictions and experimental observations.

Recently, it was demonstrated that the higher $k$ magnon pairs accessible through this interaction have long lifetimes of up to 18 $\mu$s at low temperatures \cite {rostyslav2025ultra}. Together with the control of the modes shown here by tuning the incident power and external bias field, it might unlock experimental routes for magnon quantum information and entanglement generation without the requirement of additional nonlinear elements \cite{elyasi2020resources, an2024emergent, yuan2021quantum}. Furthermore, Suhl instabilities constitute an efficient resource for the development of magnon-based stochastic computation \cite{makiuchi2021parametron, elyasi2022stochasticity, gonzalez2024spintronic}.

\textit{Author contributions--} M.A. proposed the experiment, performed measurements and data analysis, developed the Hamiltonian model, wrote the manuscript, and helped make the figures. A.B.S. performed measurements and data analysis, helped with the development of the theoretical model, made the calculations for the coupling estimation and conducted Mumax3 simulations, wrote the manuscript, and made the figures. A.B. set up and post-processed simulations on Mumax3. C.A.P contributed to the theory development and supervised the writing. Y.M.B supervised the theory and gave feedback on the writing. H.S.J.v.d.Z. provided the setup, gave feedback on data acquisition, helped devise the storyline, and supervised the writing process. G.A.S. contributed to the conception of the experiment, supervised the project and experimental work, came up with the explanation via three-wave mixing, helped devise the storyline, and gave feedback on the manuscript.

\textit{Acknowledgements--} We acknowledge V.A.S.V. Bittencourt and R. Dash for helpful discussions, and thank T. Bras for the help with the experimental setup. A. B.-S., H.S.J.vdZ. and Y.M.B. acknowledge support by the Dutch Research Council (NWO) under the project ``Ronde Open Competitie XL" (file number OCENW.XL21.XL21.058). M.A. and G.A.S. acknowledge support by the Dutch Research Council (NWO) under the project number VI.C.212.087 of the research programme VICI round 2021. C.A.P. acknowledges the support of the Novo Nordisk Foundation, NNF Quantum Computing Programme.

\bibliography{main}

\begin{thebibliography}{7}%
\makeatletter
\providecommand \@ifxundefined [1]{%
 \@ifx{#1\undefined}
}%
\providecommand \@ifnum [1]{%
 \ifnum #1\expandafter \@firstoftwo
 \else \expandafter \@secondoftwo
 \fi
}%
\providecommand \@ifx [1]{%
 \ifx #1\expandafter \@firstoftwo
 \else \expandafter \@secondoftwo
 \fi
}%
\providecommand \natexlab [1]{#1}%
\providecommand \enquote  [1]{``#1''}%
\providecommand \bibnamefont  [1]{#1}%
\providecommand \bibfnamefont [1]{#1}%
\providecommand \citenamefont [1]{#1}%
\providecommand \href@noop [0]{\@secondoftwo}%
\providecommand \href [0]{\begingroup \@sanitize@url \@href}%
\providecommand \@href[1]{\@@startlink{#1}\@@href}%
\providecommand \@@href[1]{\endgroup#1\@@endlink}%
\providecommand \@sanitize@url [0]{\catcode `\\12\catcode `\$12\catcode `\&12\catcode `\#12\catcode `\^12\catcode `\_12\catcode `\%12\relax}%
\providecommand \@@startlink[1]{}%
\providecommand \@@endlink[0]{}%
\providecommand \url  [0]{\begingroup\@sanitize@url \@url }%
\providecommand \@url [1]{\endgroup\@href {#1}{\urlprefix }}%
\providecommand \urlprefix  [0]{URL }%
\providecommand \Eprint [0]{\href }%
\providecommand \doibase [0]{https://doi.org/}%
\providecommand \selectlanguage [0]{\@gobble}%
\providecommand \bibinfo  [0]{\@secondoftwo}%
\providecommand \bibfield  [0]{\@secondoftwo}%
\providecommand \translation [1]{[#1]}%
\providecommand \BibitemOpen [0]{}%
\providecommand \bibitemStop [0]{}%
\providecommand \bibitemNoStop [0]{.\EOS\space}%
\providecommand \EOS [0]{\spacefactor3000\relax}%
\providecommand \BibitemShut  [1]{\csname bibitem#1\endcsname}%
\let\auto@bib@innerbib\@empty
%</preamble>
\bibitem [{\citenamefont {Vansteenkiste}\ \emph {et~al.}(2014)\citenamefont {Vansteenkiste}, \citenamefont {Leliaert}, \citenamefont {Dvornik}, \citenamefont {Helsen}, \citenamefont {Garcia-Sanchez},\ and\ \citenamefont {{Van Waeyenberge}}}]{Vansteenkiste2014}%
  \BibitemOpen
  \bibfield  {author} {\bibinfo {author} {\bibfnamefont {A.}~\bibnamefont {Vansteenkiste}}, \bibinfo {author} {\bibfnamefont {J.}~\bibnamefont {Leliaert}}, \bibinfo {author} {\bibfnamefont {M.}~\bibnamefont {Dvornik}}, \bibinfo {author} {\bibfnamefont {M.}~\bibnamefont {Helsen}}, \bibinfo {author} {\bibfnamefont {F.}~\bibnamefont {Garcia-Sanchez}},\ and\ \bibinfo {author} {\bibfnamefont {B.}~\bibnamefont {{Van Waeyenberge}}},\ }\bibfield  {title} {\bibinfo {title} {{The design and verification of Mumax3}},\ }\href {https://doi.org/10.1063/1.4899186} {\bibfield  {journal} {\bibinfo  {journal} {AIP Advances}\ }\textbf {\bibinfo {volume} {4}},\ \bibinfo {pages} {107133} (\bibinfo {year} {2014})}\BibitemShut {NoStop}%
\bibitem [{\citenamefont {Exl}\ \emph {et~al.}(2014)\citenamefont {Exl}, \citenamefont {Bance}, \citenamefont {Reichel}, \citenamefont {Schrefl}, \citenamefont {{Peter Stimming}},\ and\ \citenamefont {Mauser}}]{Exl2014}%
  \BibitemOpen
  \bibfield  {author} {\bibinfo {author} {\bibfnamefont {L.}~\bibnamefont {Exl}}, \bibinfo {author} {\bibfnamefont {S.}~\bibnamefont {Bance}}, \bibinfo {author} {\bibfnamefont {F.}~\bibnamefont {Reichel}}, \bibinfo {author} {\bibfnamefont {T.}~\bibnamefont {Schrefl}}, \bibinfo {author} {\bibfnamefont {H.}~\bibnamefont {{Peter Stimming}}},\ and\ \bibinfo {author} {\bibfnamefont {N.~J.}\ \bibnamefont {Mauser}},\ }\bibfield  {title} {\bibinfo {title} {{LaBonte's method revisited: An effective steepest descent method for micromagnetic energy minimization}},\ }\href {https://doi.org/10.1063/1.4862839} {\bibfield  {journal} {\bibinfo  {journal} {Journal of Applied Physics}\ }\textbf {\bibinfo {volume} {115}},\ \bibinfo {pages} {17D118} (\bibinfo {year} {2014})}\BibitemShut {NoStop}%
\bibitem [{\citenamefont {Sparks}(1970)}]{Sparks1970}%
  \BibitemOpen
  \bibfield  {author} {\bibinfo {author} {\bibfnamefont {M.}~\bibnamefont {Sparks}},\ }\bibfield  {title} {\bibinfo {title} {Ferromagnetic resonance in thin films. i. theory of normal-mode frequencies},\ }\href {https://doi.org/10.1103/PhysRevB.1.3831} {\bibfield  {journal} {\bibinfo  {journal} {Phys. Rev. B}\ }\textbf {\bibinfo {volume} {1}},\ \bibinfo {pages} {3831} (\bibinfo {year} {1970})}\BibitemShut {NoStop}%
\bibitem [{\citenamefont {L'vov}(2012)}]{l2012waveturbulence}%
  \BibitemOpen
  \bibfield  {author} {\bibinfo {author} {\bibfnamefont {V.~S.}\ \bibnamefont {L'vov}},\ }\href@noop {} {\emph {\bibinfo {title} {Wave turbulence under parametric excitation: applications to magnets}}}\ (\bibinfo  {publisher} {Springer Science \& Business Media},\ \bibinfo {year} {2012})\BibitemShut {NoStop}%
\bibitem [{\citenamefont {Suhl}(1957)}]{suhl1957theory}%
  \BibitemOpen
  \bibfield  {author} {\bibinfo {author} {\bibfnamefont {H.}~\bibnamefont {Suhl}},\ }\bibfield  {title} {\bibinfo {title} {The theory of ferromagnetic resonance at high signal powers},\ }\href {https://www.sciencedirect.com/science/article/pii/0022369757900100} {\bibfield  {journal} {\bibinfo  {journal} {J. Phys. Chem. Solids}\ }\textbf {\bibinfo {volume} {1}},\ \bibinfo {pages} {209} (\bibinfo {year} {1957})}\BibitemShut {NoStop}%
\bibitem [{\citenamefont {Zhang}\ \emph {et~al.}(2014)\citenamefont {Zhang}, \citenamefont {Zou}, \citenamefont {Jiang},\ and\ \citenamefont {Tang}}]{zhang2014strongly}%
  \BibitemOpen
  \bibfield  {author} {\bibinfo {author} {\bibfnamefont {X.}~\bibnamefont {Zhang}}, \bibinfo {author} {\bibfnamefont {C.-L.}\ \bibnamefont {Zou}}, \bibinfo {author} {\bibfnamefont {L.}~\bibnamefont {Jiang}},\ and\ \bibinfo {author} {\bibfnamefont {H.~X.}\ \bibnamefont {Tang}},\ }\bibfield  {title} {\bibinfo {title} {Strongly coupled magnons and cavity microwave photons},\ }\href@noop {} {\bibfield  {journal} {\bibinfo  {journal} {Physical Review Letters}\ }\textbf {\bibinfo {volume} {113}},\ \bibinfo {pages} {156401} (\bibinfo {year} {2014})}\BibitemShut {NoStop}%
\bibitem [{\citenamefont {Gely}\ \emph {et~al.}(2023)\citenamefont {Gely}, \citenamefont {Sanz~Mora}, \citenamefont {Yanai}, \citenamefont {Van~der Spek}, \citenamefont {Bothner},\ and\ \citenamefont {Steele}}]{gely2023apparent}%
  \BibitemOpen
  \bibfield  {author} {\bibinfo {author} {\bibfnamefont {M.~F.}\ \bibnamefont {Gely}}, \bibinfo {author} {\bibfnamefont {A.}~\bibnamefont {Sanz~Mora}}, \bibinfo {author} {\bibfnamefont {S.}~\bibnamefont {Yanai}}, \bibinfo {author} {\bibfnamefont {R.}~\bibnamefont {Van~der Spek}}, \bibinfo {author} {\bibfnamefont {D.}~\bibnamefont {Bothner}},\ and\ \bibinfo {author} {\bibfnamefont {G.~A.}\ \bibnamefont {Steele}},\ }\bibfield  {title} {\bibinfo {title} {Apparent nonlinear damping triggered by quantum fluctuations},\ }\href@noop {} {\bibfield  {journal} {\bibinfo  {journal} {Nature Communications}\ }\textbf {\bibinfo {volume} {14}},\ \bibinfo {pages} {7566} (\bibinfo {year} {2023})}\BibitemShut {NoStop}%
\end{thebibliography}%


\begin{thebibliography}{46}%
\makeatletter
\providecommand \@ifxundefined [1]{%
 \@ifx{#1\undefined}
}%
\providecommand \@ifnum [1]{%
 \ifnum #1\expandafter \@firstoftwo
 \else \expandafter \@secondoftwo
 \fi
}%
\providecommand \@ifx [1]{%
 \ifx #1\expandafter \@firstoftwo
 \else \expandafter \@secondoftwo
 \fi
}%
\providecommand \natexlab [1]{#1}%
\providecommand \enquote  [1]{``#1''}%
\providecommand \bibnamefont  [1]{#1}%
\providecommand \bibfnamefont [1]{#1}%
\providecommand \citenamefont [1]{#1}%
\providecommand \href@noop [0]{\@secondoftwo}%
\providecommand \href [0]{\begingroup \@sanitize@url \@href}%
\providecommand \@href[1]{\@@startlink{#1}\@@href}%
\providecommand \@@href[1]{\endgroup#1\@@endlink}%
\providecommand \@sanitize@url [0]{\catcode `\\12\catcode `\$12\catcode `\&12\catcode `\#12\catcode `\^12\catcode `\_12\catcode `\%12\relax}%
\providecommand \@@startlink[1]{}%
\providecommand \@@endlink[0]{}%
\providecommand \url  [0]{\begingroup\@sanitize@url \@url }%
\providecommand \@url [1]{\endgroup\@href {#1}{\urlprefix }}%
\providecommand \urlprefix  [0]{URL }%
\providecommand \Eprint [0]{\href }%
\providecommand \doibase [0]{https://doi.org/}%
\providecommand \selectlanguage [0]{\@gobble}%
\providecommand \bibinfo  [0]{\@secondoftwo}%
\providecommand \bibfield  [0]{\@secondoftwo}%
\providecommand \translation [1]{[#1]}%
\providecommand \BibitemOpen [0]{}%
\providecommand \bibitemStop [0]{}%
\providecommand \bibitemNoStop [0]{.\EOS\space}%
\providecommand \EOS [0]{\spacefactor3000\relax}%
\providecommand \BibitemShut  [1]{\csname bibitem#1\endcsname}%
\let\auto@bib@innerbib\@empty
%</preamble>
\bibitem [{\citenamefont {Griffiths}(1946)}]{griffiths1946anomalous}%
  \BibitemOpen
  \bibfield  {author} {\bibinfo {author} {\bibfnamefont {J.~H.}\ \bibnamefont {Griffiths}},\ }\bibfield  {title} {\bibinfo {title} {Anomalous high-frequency resistance of ferromagnetic metals},\ }\href {https://www.nature.com/articles/158670a0} {\bibfield  {journal} {\bibinfo  {journal} {Nature}\ }\textbf {\bibinfo {volume} {158}},\ \bibinfo {pages} {670} (\bibinfo {year} {1946})}\BibitemShut {NoStop}%
\bibitem [{\citenamefont {Kittel}(1948)}]{kittel1948theory}%
  \BibitemOpen
  \bibfield  {author} {\bibinfo {author} {\bibfnamefont {C.}~\bibnamefont {Kittel}},\ }\bibfield  {title} {\bibinfo {title} {On the theory of ferromagnetic resonance absorption},\ }\href {https://journals.aps.org/pr/abstract/10.1103/PhysRev.73.155} {\bibfield  {journal} {\bibinfo  {journal} {Phys. Rev.}\ }\textbf {\bibinfo {volume} {73}},\ \bibinfo {pages} {155} (\bibinfo {year} {1948})}\BibitemShut {NoStop}%
\bibitem [{\citenamefont {Kruglyak}\ \emph {et~al.}(2010)\citenamefont {Kruglyak}, \citenamefont {Demokritov},\ and\ \citenamefont {Grundler}}]{kruglyak2010magnonics}%
  \BibitemOpen
  \bibfield  {author} {\bibinfo {author} {\bibfnamefont {V.}~\bibnamefont {Kruglyak}}, \bibinfo {author} {\bibfnamefont {S.}~\bibnamefont {Demokritov}},\ and\ \bibinfo {author} {\bibfnamefont {D.}~\bibnamefont {Grundler}},\ }\bibfield  {title} {\bibinfo {title} {Magnonics},\ }\href {https://iopscience.iop.org/article/10.1088/0022-3727/43/26/264001} {\bibfield  {journal} {\bibinfo  {journal} {J. Phys. D: Appl.}\ }\textbf {\bibinfo {volume} {43}},\ \bibinfo {pages} {264001} (\bibinfo {year} {2010})}\BibitemShut {NoStop}%
\bibitem [{\citenamefont {{Zare Rameshti}}\ \emph {et~al.}(2022)\citenamefont {{Zare Rameshti}}, \citenamefont {{Viola Kusminskiy}}, \citenamefont {Haigh}, \citenamefont {Usami}, \citenamefont {Lachance-Quirion}, \citenamefont {Nakamura}, \citenamefont {Hu}, \citenamefont {Tang}, \citenamefont {Bauer},\ and\ \citenamefont {Blanter}}]{Zare2022cavity}%
  \BibitemOpen
  \bibfield  {author} {\bibinfo {author} {\bibfnamefont {B.}~\bibnamefont {{Zare Rameshti}}}, \bibinfo {author} {\bibfnamefont {S.}~\bibnamefont {{Viola Kusminskiy}}}, \bibinfo {author} {\bibfnamefont {J.~A.}\ \bibnamefont {Haigh}}, \bibinfo {author} {\bibfnamefont {K.}~\bibnamefont {Usami}}, \bibinfo {author} {\bibfnamefont {D.}~\bibnamefont {Lachance-Quirion}}, \bibinfo {author} {\bibfnamefont {Y.}~\bibnamefont {Nakamura}}, \bibinfo {author} {\bibfnamefont {C.-M.}\ \bibnamefont {Hu}}, \bibinfo {author} {\bibfnamefont {H.~X.}\ \bibnamefont {Tang}}, \bibinfo {author} {\bibfnamefont {G.~E.}\ \bibnamefont {Bauer}},\ and\ \bibinfo {author} {\bibfnamefont {Y.~M.}\ \bibnamefont {Blanter}},\ }\bibfield  {title} {\bibinfo {title} {Cavity magnonics},\ }\href {https://doi.org/https://doi.org/10.1016/j.physrep.2022.06.001} {\bibfield  {journal} {\bibinfo  {journal} {Phys. Rep.}\ }\textbf {\bibinfo {volume} {979}},\ \bibinfo {pages} {1} (\bibinfo {year} {2022})}\BibitemShut {NoStop}%
\bibitem [{\citenamefont {Yuan}\ \emph {et~al.}(2022)\citenamefont {Yuan}, \citenamefont {Cao}, \citenamefont {Kamra}, \citenamefont {Duine},\ and\ \citenamefont {Yan}}]{yuan2021quantum}%
  \BibitemOpen
  \bibfield  {author} {\bibinfo {author} {\bibfnamefont {H.}~\bibnamefont {Yuan}}, \bibinfo {author} {\bibfnamefont {Y.}~\bibnamefont {Cao}}, \bibinfo {author} {\bibfnamefont {A.}~\bibnamefont {Kamra}}, \bibinfo {author} {\bibfnamefont {R.~A.}\ \bibnamefont {Duine}},\ and\ \bibinfo {author} {\bibfnamefont {P.}~\bibnamefont {Yan}},\ }\bibfield  {title} {\bibinfo {title} {Quantum magnonics: When magnon spintronics meets quantum information science},\ }\href {https://doi.org/https://doi.org/10.1016/j.physrep.2022.03.002} {\bibfield  {journal} {\bibinfo  {journal} {Phys. Rep.}\ }\textbf {\bibinfo {volume} {965}},\ \bibinfo {pages} {1} (\bibinfo {year} {2022})}\BibitemShut {NoStop}%
\bibitem [{\citenamefont {Pirro}\ \emph {et~al.}(2021)\citenamefont {Pirro}, \citenamefont {Vasyuchka}, \citenamefont {Serga},\ and\ \citenamefont {Hillebrands}}]{pirro2021advances}%
  \BibitemOpen
  \bibfield  {author} {\bibinfo {author} {\bibfnamefont {P.}~\bibnamefont {Pirro}}, \bibinfo {author} {\bibfnamefont {V.~I.}\ \bibnamefont {Vasyuchka}}, \bibinfo {author} {\bibfnamefont {A.~A.}\ \bibnamefont {Serga}},\ and\ \bibinfo {author} {\bibfnamefont {B.}~\bibnamefont {Hillebrands}},\ }\bibfield  {title} {\bibinfo {title} {Advances in coherent magnonics},\ }\href {https://www.nature.com/articles/s41578-021-00332-w} {\bibfield  {journal} {\bibinfo  {journal} {Nat. Rev. Mater.}\ }\textbf {\bibinfo {volume} {6}},\ \bibinfo {pages} {1114} (\bibinfo {year} {2021})}\BibitemShut {NoStop}%
\bibitem [{\citenamefont {Tabuchi}\ \emph {et~al.}(2015)\citenamefont {Tabuchi}, \citenamefont {Ishino}, \citenamefont {Noguchi}, \citenamefont {Ishikawa}, \citenamefont {Yamazaki}, \citenamefont {Usami},\ and\ \citenamefont {Nakamura}}]{tabuchi2015coherent}%
  \BibitemOpen
  \bibfield  {author} {\bibinfo {author} {\bibfnamefont {Y.}~\bibnamefont {Tabuchi}}, \bibinfo {author} {\bibfnamefont {S.}~\bibnamefont {Ishino}}, \bibinfo {author} {\bibfnamefont {A.}~\bibnamefont {Noguchi}}, \bibinfo {author} {\bibfnamefont {T.}~\bibnamefont {Ishikawa}}, \bibinfo {author} {\bibfnamefont {R.}~\bibnamefont {Yamazaki}}, \bibinfo {author} {\bibfnamefont {K.}~\bibnamefont {Usami}},\ and\ \bibinfo {author} {\bibfnamefont {Y.}~\bibnamefont {Nakamura}},\ }\bibfield  {title} {\bibinfo {title} {Coherent coupling between a ferromagnetic magnon and a superconducting qubit},\ }\href {https://www.science.org/doi/10.1126/science.aaa3693} {\bibfield  {journal} {\bibinfo  {journal} {Science}\ }\textbf {\bibinfo {volume} {349}},\ \bibinfo {pages} {405} (\bibinfo {year} {2015})}\BibitemShut {NoStop}%
\bibitem [{\citenamefont {Song}\ \emph {et~al.}(2025)\citenamefont {Song}, \citenamefont {Polakovic}, \citenamefont {Lim}, \citenamefont {Cecil}, \citenamefont {Pearson}, \citenamefont {Divan}, \citenamefont {Kwok}, \citenamefont {Welp}, \citenamefont {Hoffmann}, \citenamefont {Kim} \emph {et~al.}}]{song2025single}%
  \BibitemOpen
  \bibfield  {author} {\bibinfo {author} {\bibfnamefont {M.}~\bibnamefont {Song}}, \bibinfo {author} {\bibfnamefont {T.}~\bibnamefont {Polakovic}}, \bibinfo {author} {\bibfnamefont {J.}~\bibnamefont {Lim}}, \bibinfo {author} {\bibfnamefont {T.~W.}\ \bibnamefont {Cecil}}, \bibinfo {author} {\bibfnamefont {J.}~\bibnamefont {Pearson}}, \bibinfo {author} {\bibfnamefont {R.}~\bibnamefont {Divan}}, \bibinfo {author} {\bibfnamefont {W.-K.}\ \bibnamefont {Kwok}}, \bibinfo {author} {\bibfnamefont {U.}~\bibnamefont {Welp}}, \bibinfo {author} {\bibfnamefont {A.}~\bibnamefont {Hoffmann}}, \bibinfo {author} {\bibfnamefont {K.-J.}\ \bibnamefont {Kim}}, \emph {et~al.},\ }\bibfield  {title} {\bibinfo {title} {Single-shot magnon interference in a magnon-superconducting-resonator hybrid circuit},\ }\href {https://www.nature.com/articles/s41467-025-58482-2} {\bibfield  {journal} {\bibinfo  {journal} {Nature Comms.}\ }\textbf {\bibinfo {volume} {16}},\ \bibinfo {pages} {3649} (\bibinfo {year} {2025})}\BibitemShut {NoStop}%
\bibitem [{\citenamefont {Li}\ \emph {et~al.}(2020)\citenamefont {Li}, \citenamefont {Zhang}, \citenamefont {Tyberkevych}, \citenamefont {Kwok}, \citenamefont {Hoffmann},\ and\ \citenamefont {Novosad}}]{li2020hybrid}%
  \BibitemOpen
  \bibfield  {author} {\bibinfo {author} {\bibfnamefont {Y.}~\bibnamefont {Li}}, \bibinfo {author} {\bibfnamefont {W.}~\bibnamefont {Zhang}}, \bibinfo {author} {\bibfnamefont {V.}~\bibnamefont {Tyberkevych}}, \bibinfo {author} {\bibfnamefont {W.-K.}\ \bibnamefont {Kwok}}, \bibinfo {author} {\bibfnamefont {A.}~\bibnamefont {Hoffmann}},\ and\ \bibinfo {author} {\bibfnamefont {V.}~\bibnamefont {Novosad}},\ }\bibfield  {title} {\bibinfo {title} {Hybrid magnonics: Physics, circuits, and applications for coherent information processing},\ }\href {https://pubs.aip.org/aip/jap/article/128/13/130902/287124/Hybrid-magnonics-Physics-circuits-and-applications} {\bibfield  {journal} {\bibinfo  {journal} {J. Appl. Phys.}\ }\textbf {\bibinfo {volume} {128}} (\bibinfo {year} {2020})}\BibitemShut {NoStop}%
\bibitem [{\citenamefont {Xu}\ \emph {et~al.}(2023)\citenamefont {Xu}, \citenamefont {Gu}, \citenamefont {Li}, \citenamefont {Weng}, \citenamefont {Wang}, \citenamefont {Li}, \citenamefont {Wang}, \citenamefont {Zhu},\ and\ \citenamefont {You}}]{xu2023quantum}%
  \BibitemOpen
  \bibfield  {author} {\bibinfo {author} {\bibfnamefont {D.}~\bibnamefont {Xu}}, \bibinfo {author} {\bibfnamefont {X.-K.}\ \bibnamefont {Gu}}, \bibinfo {author} {\bibfnamefont {H.-K.}\ \bibnamefont {Li}}, \bibinfo {author} {\bibfnamefont {Y.-C.}\ \bibnamefont {Weng}}, \bibinfo {author} {\bibfnamefont {Y.-P.}\ \bibnamefont {Wang}}, \bibinfo {author} {\bibfnamefont {J.}~\bibnamefont {Li}}, \bibinfo {author} {\bibfnamefont {H.}~\bibnamefont {Wang}}, \bibinfo {author} {\bibfnamefont {S.-Y.}\ \bibnamefont {Zhu}},\ and\ \bibinfo {author} {\bibfnamefont {J.}~\bibnamefont {You}},\ }\bibfield  {title} {\bibinfo {title} {Quantum control of a single magnon in a macroscopic spin system},\ }\href {https://journals.aps.org/prl/abstract/10.1103/PhysRevLett.130.193603} {\bibfield  {journal} {\bibinfo  {journal} {Phys. Rev. Lett.}\ }\textbf {\bibinfo {volume} {130}},\ \bibinfo {pages} {193603} (\bibinfo {year} {2023})}\BibitemShut {NoStop}%
\bibitem [{\citenamefont {Chumak}\ \emph {et~al.}(2014)\citenamefont {Chumak}, \citenamefont {Serga},\ and\ \citenamefont {Hillebrands}}]{chumak2014magnon_transistor}%
  \BibitemOpen
  \bibfield  {author} {\bibinfo {author} {\bibfnamefont {A.~V.}\ \bibnamefont {Chumak}}, \bibinfo {author} {\bibfnamefont {A.~A.}\ \bibnamefont {Serga}},\ and\ \bibinfo {author} {\bibfnamefont {B.}~\bibnamefont {Hillebrands}},\ }\bibfield  {title} {\bibinfo {title} {Magnon transistor for all-magnon data processing},\ }\href {https://www.nature.com/articles/ncomms5700} {\bibfield  {journal} {\bibinfo  {journal} {Nat. comms.}\ }\textbf {\bibinfo {volume} {5}},\ \bibinfo {pages} {4700} (\bibinfo {year} {2014})}\BibitemShut {NoStop}%
\bibitem [{\citenamefont {Soykal}\ and\ \citenamefont {Flatt{\'e}}(2010)}]{soykal2010strong}%
  \BibitemOpen
  \bibfield  {author} {\bibinfo {author} {\bibfnamefont {{\"O}.~O.}\ \bibnamefont {Soykal}}\ and\ \bibinfo {author} {\bibfnamefont {M.}~\bibnamefont {Flatt{\'e}}},\ }\bibfield  {title} {\bibinfo {title} {Strong field interactions between a nanomagnet and a photonic cavity},\ }\href@noop {} {\bibfield  {journal} {\bibinfo  {journal} {Physical Review Letters}\ }\textbf {\bibinfo {volume} {104}},\ \bibinfo {pages} {077202} (\bibinfo {year} {2010})}\BibitemShut {NoStop}%
\bibitem [{\citenamefont {Huebl}\ \emph {et~al.}(2013)\citenamefont {Huebl}, \citenamefont {Zollitsch}, \citenamefont {Lotze}, \citenamefont {Hocke}, \citenamefont {Greifenstein}, \citenamefont {Marx}, \citenamefont {Gross},\ and\ \citenamefont {Goennenwein}}]{huebl2013high}%
  \BibitemOpen
  \bibfield  {author} {\bibinfo {author} {\bibfnamefont {H.}~\bibnamefont {Huebl}}, \bibinfo {author} {\bibfnamefont {C.~W.}\ \bibnamefont {Zollitsch}}, \bibinfo {author} {\bibfnamefont {J.}~\bibnamefont {Lotze}}, \bibinfo {author} {\bibfnamefont {F.}~\bibnamefont {Hocke}}, \bibinfo {author} {\bibfnamefont {M.}~\bibnamefont {Greifenstein}}, \bibinfo {author} {\bibfnamefont {A.}~\bibnamefont {Marx}}, \bibinfo {author} {\bibfnamefont {R.}~\bibnamefont {Gross}},\ and\ \bibinfo {author} {\bibfnamefont {S.~T.}\ \bibnamefont {Goennenwein}},\ }\bibfield  {title} {\bibinfo {title} {High cooperativity in coupled microwave resonator ferrimagnetic insulator hybrids},\ }\href@noop {} {\bibfield  {journal} {\bibinfo  {journal} {Physical Review Letters}\ }\textbf {\bibinfo {volume} {111}},\ \bibinfo {pages} {127003} (\bibinfo {year} {2013})}\BibitemShut {NoStop}%
\bibitem [{\citenamefont {Zhang}\ \emph {et~al.}(2016)\citenamefont {Zhang}, \citenamefont {Zou}, \citenamefont {Jiang},\ and\ \citenamefont {Tang}}]{zhang2016cavity_magnomech}%
  \BibitemOpen
  \bibfield  {author} {\bibinfo {author} {\bibfnamefont {X.}~\bibnamefont {Zhang}}, \bibinfo {author} {\bibfnamefont {C.-L.}\ \bibnamefont {Zou}}, \bibinfo {author} {\bibfnamefont {L.}~\bibnamefont {Jiang}},\ and\ \bibinfo {author} {\bibfnamefont {H.~X.}\ \bibnamefont {Tang}},\ }\bibfield  {title} {\bibinfo {title} {Cavity magnomechanics},\ }\href {https://www.science.org/doi/10.1126/sciadv.1501286} {\bibfield  {journal} {\bibinfo  {journal} {Science adv.}\ }\textbf {\bibinfo {volume} {2}},\ \bibinfo {pages} {e1501286} (\bibinfo {year} {2016})}\BibitemShut {NoStop}%
\bibitem [{\citenamefont {Potts}\ \emph {et~al.}(2021)\citenamefont {Potts}, \citenamefont {Varga}, \citenamefont {Bittencourt}, \citenamefont {Kusminskiy},\ and\ \citenamefont {Davis}}]{potts2021dynamical}%
  \BibitemOpen
  \bibfield  {author} {\bibinfo {author} {\bibfnamefont {C.~A.}\ \bibnamefont {Potts}}, \bibinfo {author} {\bibfnamefont {E.}~\bibnamefont {Varga}}, \bibinfo {author} {\bibfnamefont {V.~A. S.~V.}\ \bibnamefont {Bittencourt}}, \bibinfo {author} {\bibfnamefont {S.~V.}\ \bibnamefont {Kusminskiy}},\ and\ \bibinfo {author} {\bibfnamefont {J.~P.}\ \bibnamefont {Davis}},\ }\bibfield  {title} {\bibinfo {title} {Dynamical backaction magnomechanics},\ }\href {https://doi.org/10.1103/PhysRevX.11.031053} {\bibfield  {journal} {\bibinfo  {journal} {Phys. Rev. X}\ }\textbf {\bibinfo {volume} {11}},\ \bibinfo {pages} {031053} (\bibinfo {year} {2021})}\BibitemShut {NoStop}%
\bibitem [{\citenamefont {Bader}\ and\ \citenamefont {Parkin}(2010)}]{bader2010spintronics}%
  \BibitemOpen
  \bibfield  {author} {\bibinfo {author} {\bibfnamefont {S.~D.}\ \bibnamefont {Bader}}\ and\ \bibinfo {author} {\bibfnamefont {S.~S.~P.}\ \bibnamefont {Parkin}},\ }\bibfield  {title} {\bibinfo {title} {Spintronics},\ }\href {https://www.annualreviews.org/content/journals/10.1146/annurev-conmatphys-070909-104123} {\bibfield  {journal} {\bibinfo  {journal} {Annu. Rev. Condens. Matter Phys.}\ }\textbf {\bibinfo {volume} {1}},\ \bibinfo {pages} {71} (\bibinfo {year} {2010})}\BibitemShut {NoStop}%
\bibitem [{\citenamefont {Lachance-Quirion}\ \emph {et~al.}(2017)\citenamefont {Lachance-Quirion}, \citenamefont {Tabuchi}, \citenamefont {Ishino}, \citenamefont {Noguchi}, \citenamefont {Ishikawa}, \citenamefont {Yamazaki},\ and\ \citenamefont {Nakamura}}]{lachance2017resolving}%
  \BibitemOpen
  \bibfield  {author} {\bibinfo {author} {\bibfnamefont {D.}~\bibnamefont {Lachance-Quirion}}, \bibinfo {author} {\bibfnamefont {Y.}~\bibnamefont {Tabuchi}}, \bibinfo {author} {\bibfnamefont {S.}~\bibnamefont {Ishino}}, \bibinfo {author} {\bibfnamefont {A.}~\bibnamefont {Noguchi}}, \bibinfo {author} {\bibfnamefont {T.}~\bibnamefont {Ishikawa}}, \bibinfo {author} {\bibfnamefont {R.}~\bibnamefont {Yamazaki}},\ and\ \bibinfo {author} {\bibfnamefont {Y.}~\bibnamefont {Nakamura}},\ }\bibfield  {title} {\bibinfo {title} {Resolving quanta of collective spin excitations in a millimeter-sized ferromagnet},\ }\href {https://doi.org/10.1126/sciadv.1603150} {\bibfield  {journal} {\bibinfo  {journal} {Science Adv.}\ }\textbf {\bibinfo {volume} {3}},\ \bibinfo {pages} {e1603150} (\bibinfo {year} {2017})}\BibitemShut {NoStop}%
\bibitem [{\citenamefont {Rao}\ \emph {et~al.}(2024)\citenamefont {Rao}, \citenamefont {Meng}, \citenamefont {Han}, \citenamefont {Zhu}, \citenamefont {Ding},\ and\ \citenamefont {An}}]{rao2024braiding}%
  \BibitemOpen
  \bibfield  {author} {\bibinfo {author} {\bibfnamefont {Z.}~\bibnamefont {Rao}}, \bibinfo {author} {\bibfnamefont {C.}~\bibnamefont {Meng}}, \bibinfo {author} {\bibfnamefont {Y.}~\bibnamefont {Han}}, \bibinfo {author} {\bibfnamefont {L.}~\bibnamefont {Zhu}}, \bibinfo {author} {\bibfnamefont {K.}~\bibnamefont {Ding}},\ and\ \bibinfo {author} {\bibfnamefont {Z.}~\bibnamefont {An}},\ }\bibfield  {title} {\bibinfo {title} {Braiding reflectionless states in non-hermitian magnonics},\ }\href {https://doi.org/10.1038/s41567-024-02667-x} {\bibfield  {journal} {\bibinfo  {journal} {Nat. Phys.}\ }\textbf {\bibinfo {volume} {20}},\ \bibinfo {pages} {1904} (\bibinfo {year} {2024})}\BibitemShut {NoStop}%
\bibitem [{\citenamefont {Kamra}\ and\ \citenamefont {Belzig}(2016)}]{kamra2016super}%
  \BibitemOpen
  \bibfield  {author} {\bibinfo {author} {\bibfnamefont {A.}~\bibnamefont {Kamra}}\ and\ \bibinfo {author} {\bibfnamefont {W.}~\bibnamefont {Belzig}},\ }\bibfield  {title} {\bibinfo {title} {Super-poissonian shot noise of squeezed-magnon mediated spin transport},\ }\href {https://doi.org/10.1103/PhysRevLett.116.146601} {\bibfield  {journal} {\bibinfo  {journal} {Phys. Rev. Lett.}\ }\textbf {\bibinfo {volume} {116}},\ \bibinfo {pages} {146601} (\bibinfo {year} {2016})}\BibitemShut {NoStop}%
\bibitem [{\citenamefont {Kamra}\ \emph {et~al.}(2020)\citenamefont {Kamra}, \citenamefont {Belzig},\ and\ \citenamefont {Brataas}}]{kamra2020magnon}%
  \BibitemOpen
  \bibfield  {author} {\bibinfo {author} {\bibfnamefont {A.}~\bibnamefont {Kamra}}, \bibinfo {author} {\bibfnamefont {W.}~\bibnamefont {Belzig}},\ and\ \bibinfo {author} {\bibfnamefont {A.}~\bibnamefont {Brataas}},\ }\bibfield  {title} {\bibinfo {title} {Magnon-squeezing as a niche of quantum magnonics},\ }\href {https://doi.org/10.1063/5.0021099} {\bibfield  {journal} {\bibinfo  {journal} {Appl. Phys. Lett.}\ }\textbf {\bibinfo {volume} {117}},\ \bibinfo {pages} {090501} (\bibinfo {year} {2020})}\BibitemShut {NoStop}%
\bibitem [{\citenamefont {Elyasi}\ \emph {et~al.}(2020)\citenamefont {Elyasi}, \citenamefont {Blanter},\ and\ \citenamefont {Bauer}}]{elyasi2020resources}%
  \BibitemOpen
  \bibfield  {author} {\bibinfo {author} {\bibfnamefont {M.}~\bibnamefont {Elyasi}}, \bibinfo {author} {\bibfnamefont {Y.~M.}\ \bibnamefont {Blanter}},\ and\ \bibinfo {author} {\bibfnamefont {G.~E.}\ \bibnamefont {Bauer}},\ }\bibfield  {title} {\bibinfo {title} {Resources of nonlinear cavity magnonics for quantum information},\ }\href {https://journals.aps.org/prb/abstract/10.1103/PhysRevB.101.054402} {\bibfield  {journal} {\bibinfo  {journal} {Phys. Rev. B}\ }\textbf {\bibinfo {volume} {101}},\ \bibinfo {pages} {054402} (\bibinfo {year} {2020})}\BibitemShut {NoStop}%
\bibitem [{\citenamefont {Zhang}\ \emph {et~al.}(2019)\citenamefont {Zhang}, \citenamefont {Scully},\ and\ \citenamefont {Agarwal}}]{zhang2019quantum}%
  \BibitemOpen
  \bibfield  {author} {\bibinfo {author} {\bibfnamefont {Z.}~\bibnamefont {Zhang}}, \bibinfo {author} {\bibfnamefont {M.~O.}\ \bibnamefont {Scully}},\ and\ \bibinfo {author} {\bibfnamefont {G.~S.}\ \bibnamefont {Agarwal}},\ }\bibfield  {title} {\bibinfo {title} {Quantum entanglement between two magnon modes via {Kerr} nonlinearity driven far from equilibrium},\ }\href {https://doi.org/10.1103/PhysRevResearch.1.023021} {\bibfield  {journal} {\bibinfo  {journal} {Phys. Rev. Res.}\ }\textbf {\bibinfo {volume} {1}},\ \bibinfo {pages} {023021} (\bibinfo {year} {2019})}\BibitemShut {NoStop}%
\bibitem [{\citenamefont {Pal}\ \emph {et~al.}(2024)\citenamefont {Pal}, \citenamefont {Mondal},\ and\ \citenamefont {Barman}}]{pal2024using_magn_quantum_tech}%
  \BibitemOpen
  \bibfield  {author} {\bibinfo {author} {\bibfnamefont {P.~K.}\ \bibnamefont {Pal}}, \bibinfo {author} {\bibfnamefont {A.~K.}\ \bibnamefont {Mondal}},\ and\ \bibinfo {author} {\bibfnamefont {A.}~\bibnamefont {Barman}},\ }\bibfield  {title} {\bibinfo {title} {Using magnons as a quantum technology platform: a perspective},\ }\href {https://iopscience.iop.org/article/10.1088/1361-648X/ad6828} {\bibfield  {journal} {\bibinfo  {journal} {J. Phys.: Condens. Matter}\ }\textbf {\bibinfo {volume} {36}},\ \bibinfo {pages} {441502} (\bibinfo {year} {2024})}\BibitemShut {NoStop}%
\bibitem [{\citenamefont {Suhl}(1957)}]{suhl1957theory}%
  \BibitemOpen
  \bibfield  {author} {\bibinfo {author} {\bibfnamefont {H.}~\bibnamefont {Suhl}},\ }\bibfield  {title} {\bibinfo {title} {The theory of ferromagnetic resonance at high signal powers},\ }\href {https://www.sciencedirect.com/science/article/pii/0022369757900100} {\bibfield  {journal} {\bibinfo  {journal} {J. Phys. Chem. Solids}\ }\textbf {\bibinfo {volume} {1}},\ \bibinfo {pages} {209} (\bibinfo {year} {1957})}\BibitemShut {NoStop}%
\bibitem [{\citenamefont {Pecora}(1988)}]{pecora1988derivation}%
  \BibitemOpen
  \bibfield  {author} {\bibinfo {author} {\bibfnamefont {L.~M.}\ \bibnamefont {Pecora}},\ }\bibfield  {title} {\bibinfo {title} {Derivation and generalization of the {Suhl} spin-wave instability relations},\ }\href {https://journals.aps.org/prb/abstract/10.1103/PhysRevB.37.5473} {\bibfield  {journal} {\bibinfo  {journal} {Phys. Rev. B}\ }\textbf {\bibinfo {volume} {37}},\ \bibinfo {pages} {5473} (\bibinfo {year} {1988})}\BibitemShut {NoStop}%
\bibitem [{\citenamefont {Stancil}\ and\ \citenamefont {Prabhakar}(2009)}]{stancil2009spin}%
  \BibitemOpen
  \bibfield  {author} {\bibinfo {author} {\bibfnamefont {D.~D.}\ \bibnamefont {Stancil}}\ and\ \bibinfo {author} {\bibfnamefont {A.}~\bibnamefont {Prabhakar}},\ }\href@noop {} {\emph {\bibinfo {title} {Spin waves}}}\ (\bibinfo  {publisher} {Springer},\ \bibinfo {year} {2009})\BibitemShut {NoStop}%
\bibitem [{\citenamefont {Krivosik}\ and\ \citenamefont {Patton}(2010)}]{krivosik2010hamiltonian}%
  \BibitemOpen
  \bibfield  {author} {\bibinfo {author} {\bibfnamefont {P.}~\bibnamefont {Krivosik}}\ and\ \bibinfo {author} {\bibfnamefont {C.~E.}\ \bibnamefont {Patton}},\ }\bibfield  {title} {\bibinfo {title} {Hamiltonian formulation of nonlinear spin-wave dynamics: Theory and applications},\ }\href {https://journals.aps.org/prb/abstract/10.1103/PhysRevB.82.184428} {\bibfield  {journal} {\bibinfo  {journal} {Phys. Rev. B}\ }\textbf {\bibinfo {volume} {82}},\ \bibinfo {pages} {184428} (\bibinfo {year} {2010})}\BibitemShut {NoStop}%
\bibitem [{\citenamefont {L'vov}(2012)}]{l2012waveturbulence}%
  \BibitemOpen
  \bibfield  {author} {\bibinfo {author} {\bibfnamefont {V.~S.}\ \bibnamefont {L'vov}},\ }\href@noop {} {\emph {\bibinfo {title} {Wave turbulence under parametric excitation: applications to magnets}}}\ (\bibinfo  {publisher} {Springer Science \& Business Media},\ \bibinfo {year} {2012})\BibitemShut {NoStop}%
\bibitem [{\citenamefont {Schl{\"o}mann}\ \emph {et~al.}(1960)\citenamefont {Schl{\"o}mann}, \citenamefont {Green},\ and\ \citenamefont {Milano}}]{schlomann1960recent}%
  \BibitemOpen
  \bibfield  {author} {\bibinfo {author} {\bibfnamefont {E.}~\bibnamefont {Schl{\"o}mann}}, \bibinfo {author} {\bibfnamefont {J.}~\bibnamefont {Green}},\ and\ \bibinfo {author} {\bibfnamefont {U.}~\bibnamefont {Milano}},\ }\bibfield  {title} {\bibinfo {title} {Recent developments in ferromagnetic resonance at high power levels},\ }\href {https://pubs.aip.org/aip/jap/article/31/5/S386/147680/Recent-Developments-in-Ferromagnetic-Resonance-at} {\bibfield  {journal} {\bibinfo  {journal} {J. Appl. Phys.}\ }\textbf {\bibinfo {volume} {31}},\ \bibinfo {pages} {S386} (\bibinfo {year} {1960})}\BibitemShut {NoStop}%
\bibitem [{\citenamefont {Lee}\ \emph {et~al.}(2023)\citenamefont {Lee}, \citenamefont {Yamamoto}, \citenamefont {Umeda}, \citenamefont {Zollitsch}, \citenamefont {Elyasi}, \citenamefont {Kikkawa}, \citenamefont {Saitoh}, \citenamefont {Bauer},\ and\ \citenamefont {Kurebayashi}}]{lee2023nonlinear}%
  \BibitemOpen
  \bibfield  {author} {\bibinfo {author} {\bibfnamefont {O.}~\bibnamefont {Lee}}, \bibinfo {author} {\bibfnamefont {K.}~\bibnamefont {Yamamoto}}, \bibinfo {author} {\bibfnamefont {M.}~\bibnamefont {Umeda}}, \bibinfo {author} {\bibfnamefont {C.~W.}\ \bibnamefont {Zollitsch}}, \bibinfo {author} {\bibfnamefont {M.}~\bibnamefont {Elyasi}}, \bibinfo {author} {\bibfnamefont {T.}~\bibnamefont {Kikkawa}}, \bibinfo {author} {\bibfnamefont {E.}~\bibnamefont {Saitoh}}, \bibinfo {author} {\bibfnamefont {G.~E.}\ \bibnamefont {Bauer}},\ and\ \bibinfo {author} {\bibfnamefont {H.}~\bibnamefont {Kurebayashi}},\ }\bibfield  {title} {\bibinfo {title} {Nonlinear magnon polaritons},\ }\href {https://journals.aps.org/prl/abstract/10.1103/PhysRevLett.130.046703} {\bibfield  {journal} {\bibinfo  {journal} {Phys. Rev. Lett.}\ }\textbf {\bibinfo {volume} {130}},\ \bibinfo {pages} {046703} (\bibinfo {year} {2023})}\BibitemShut {NoStop}%
\bibitem [{\citenamefont {Barsukov}\ \emph {et~al.}(2019)\citenamefont {Barsukov}, \citenamefont {Lee}, \citenamefont {Jara}, \citenamefont {Chen}, \citenamefont {Gon{\c{c}}alves}, \citenamefont {Sha}, \citenamefont {Katine}, \citenamefont {Arias}, \citenamefont {Ivanov},\ and\ \citenamefont {Krivorotov}}]{barsukov2019giant}%
  \BibitemOpen
  \bibfield  {author} {\bibinfo {author} {\bibfnamefont {I.}~\bibnamefont {Barsukov}}, \bibinfo {author} {\bibfnamefont {H.}~\bibnamefont {Lee}}, \bibinfo {author} {\bibfnamefont {A.}~\bibnamefont {Jara}}, \bibinfo {author} {\bibfnamefont {Y.-J.}\ \bibnamefont {Chen}}, \bibinfo {author} {\bibfnamefont {A.}~\bibnamefont {Gon{\c{c}}alves}}, \bibinfo {author} {\bibfnamefont {C.}~\bibnamefont {Sha}}, \bibinfo {author} {\bibfnamefont {J.}~\bibnamefont {Katine}}, \bibinfo {author} {\bibfnamefont {R.}~\bibnamefont {Arias}}, \bibinfo {author} {\bibfnamefont {B.}~\bibnamefont {Ivanov}},\ and\ \bibinfo {author} {\bibfnamefont {I.}~\bibnamefont {Krivorotov}},\ }\bibfield  {title} {\bibinfo {title} {Giant nonlinear damping in nanoscale ferromagnets},\ }\href {https://www.science.org/doi/10.1126/sciadv.aav6943} {\bibfield  {journal} {\bibinfo  {journal} {Science adv.}\ }\textbf {\bibinfo {volume} {5}},\ \bibinfo {pages} {eaav6943} (\bibinfo {year} {2019})}\BibitemShut {NoStop}%
\bibitem [{\citenamefont {Sheng}\ \emph {et~al.}(2023)\citenamefont {Sheng}, \citenamefont {Elyasi}, \citenamefont {Chen}, \citenamefont {He}, \citenamefont {Wang}, \citenamefont {Wang}, \citenamefont {Feng}, \citenamefont {Zhang}, \citenamefont {Medlej}, \citenamefont {Liu} \emph {et~al.}}]{sheng2023nonlocal}%
  \BibitemOpen
  \bibfield  {author} {\bibinfo {author} {\bibfnamefont {L.}~\bibnamefont {Sheng}}, \bibinfo {author} {\bibfnamefont {M.}~\bibnamefont {Elyasi}}, \bibinfo {author} {\bibfnamefont {J.}~\bibnamefont {Chen}}, \bibinfo {author} {\bibfnamefont {W.}~\bibnamefont {He}}, \bibinfo {author} {\bibfnamefont {Y.}~\bibnamefont {Wang}}, \bibinfo {author} {\bibfnamefont {H.}~\bibnamefont {Wang}}, \bibinfo {author} {\bibfnamefont {H.}~\bibnamefont {Feng}}, \bibinfo {author} {\bibfnamefont {Y.}~\bibnamefont {Zhang}}, \bibinfo {author} {\bibfnamefont {I.}~\bibnamefont {Medlej}}, \bibinfo {author} {\bibfnamefont {S.}~\bibnamefont {Liu}}, \emph {et~al.},\ }\bibfield  {title} {\bibinfo {title} {Nonlocal detection of interlayer three-magnon coupling},\ }\href {https://journals.aps.org/prl/abstract/10.1103/PhysRevLett.130.046701} {\bibfield  {journal} {\bibinfo  {journal} {Phys. Rev. Lett.}\ }\textbf {\bibinfo {volume} {130}},\ \bibinfo {pages} {046701} (\bibinfo {year} {2023})}\BibitemShut {NoStop}%
\bibitem [{\citenamefont {Kimura}\ and\ \citenamefont {Shindo}(1977)}]{kimura1977single_FZ}%
  \BibitemOpen
  \bibfield  {author} {\bibinfo {author} {\bibfnamefont {S.}~\bibnamefont {Kimura}}\ and\ \bibinfo {author} {\bibfnamefont {I.}~\bibnamefont {Shindo}},\ }\bibfield  {title} {\bibinfo {title} {Single crystal growth of yig by the floating zone method},\ }\href {https://www.sciencedirect.com/science/article/pii/0022024877900458} {\bibfield  {journal} {\bibinfo  {journal} {J. Cryst. Growth}\ }\textbf {\bibinfo {volume} {41}},\ \bibinfo {pages} {192} (\bibinfo {year} {1977})}\BibitemShut {NoStop}%
\bibitem [{\citenamefont {Dillon~Jr}(1960)}]{dillon1960magnetostatic}%
  \BibitemOpen
  \bibfield  {author} {\bibinfo {author} {\bibfnamefont {J.}~\bibnamefont {Dillon~Jr}},\ }\bibfield  {title} {\bibinfo {title} {Magnetostatic modes in disks and rods},\ }\href {https://pubs.aip.org/aip/jap/article/31/9/1605/286446/Magnetostatic-Modes-in-Disks-and-Rods} {\bibfield  {journal} {\bibinfo  {journal} {J. Appl. Phys.}\ }\textbf {\bibinfo {volume} {31}},\ \bibinfo {pages} {1605} (\bibinfo {year} {1960})}\BibitemShut {NoStop}%
\bibitem [{\citenamefont {Edwards}\ \emph {et~al.}(2013)\citenamefont {Edwards}, \citenamefont {Buchmeier}, \citenamefont {Demidov},\ and\ \citenamefont {Demokritov}}]{edwards2013magnetostatic}%
  \BibitemOpen
  \bibfield  {author} {\bibinfo {author} {\bibfnamefont {E.~R.}\ \bibnamefont {Edwards}}, \bibinfo {author} {\bibfnamefont {M.}~\bibnamefont {Buchmeier}}, \bibinfo {author} {\bibfnamefont {V.~E.}\ \bibnamefont {Demidov}},\ and\ \bibinfo {author} {\bibfnamefont {S.~O.}\ \bibnamefont {Demokritov}},\ }\bibfield  {title} {\bibinfo {title} {Magnetostatic spin-wave modes of an in-plane magnetized garnet-film disk},\ }\href {https://pubs.aip.org/aip/jap/article/113/10/103901/963674/Magnetostatic-spin-wave-modes-of-an-in-plane} {\bibfield  {journal} {\bibinfo  {journal} {J. Appl. Phys.}\ }\textbf {\bibinfo {volume} {113}} (\bibinfo {year} {2013})}\BibitemShut {NoStop}%
\bibitem [{SI_()}]{SI_supp}%
  \BibitemOpen
  \href@noop {} {}\bibinfo {note} {Supplementary Information}\BibitemShut {NoStop}%
\bibitem [{\citenamefont {Probst}\ \emph {et~al.}(2015)\citenamefont {Probst}, \citenamefont {Song}, \citenamefont {Bushev}, \citenamefont {Ustinov},\ and\ \citenamefont {Weides}}]{S21fit}%
  \BibitemOpen
  \bibfield  {author} {\bibinfo {author} {\bibfnamefont {S.}~\bibnamefont {Probst}}, \bibinfo {author} {\bibfnamefont {F.}~\bibnamefont {Song}}, \bibinfo {author} {\bibfnamefont {P.~A.}\ \bibnamefont {Bushev}}, \bibinfo {author} {\bibfnamefont {A.~V.}\ \bibnamefont {Ustinov}},\ and\ \bibinfo {author} {\bibfnamefont {M.}~\bibnamefont {Weides}},\ }\bibfield  {title} {\bibinfo {title} {Efficient and robust analysis of complex scattering data under noise in microwave resonators},\ }\href {https://pubs.aip.org/aip/rsi/article/86/2/024706/360955/Efficient-and-robust-analysis-of-complex} {\bibfield  {journal} {\bibinfo  {journal} {Rev. Sci. Instrum.}\ }\textbf {\bibinfo {volume} {86}} (\bibinfo {year} {2015})}\BibitemShut {NoStop}%
\bibitem [{\citenamefont {Suhl}\ and\ \citenamefont {Zhang}(1988)}]{suhl1988spin}%
  \BibitemOpen
  \bibfield  {author} {\bibinfo {author} {\bibfnamefont {H.}~\bibnamefont {Suhl}}\ and\ \bibinfo {author} {\bibfnamefont {X.}~\bibnamefont {Zhang}},\ }\bibfield  {title} {\bibinfo {title} {Spin-wave instabilities and their revival by nonlinear mechanics},\ }\href {https://pubs.aip.org/aip/jap/article/63/8/4147/501782/Spin-wave-instabilities-and-their-revival-by} {\bibfield  {journal} {\bibinfo  {journal} {J. Appl. Phys.}\ }\textbf {\bibinfo {volume} {63}},\ \bibinfo {pages} {4147} (\bibinfo {year} {1988})}\BibitemShut {NoStop}%
\bibitem [{\citenamefont {Hill}\ \emph {et~al.}(1978)\citenamefont {Hill}, \citenamefont {Johnson}, \citenamefont {Kawasaki},\ and\ \citenamefont {MacDonald}}]{hill1978cw_3waveoptics}%
  \BibitemOpen
  \bibfield  {author} {\bibinfo {author} {\bibfnamefont {K.}~\bibnamefont {Hill}}, \bibinfo {author} {\bibfnamefont {D.}~\bibnamefont {Johnson}}, \bibinfo {author} {\bibfnamefont {B.}~\bibnamefont {Kawasaki}},\ and\ \bibinfo {author} {\bibfnamefont {R.}~\bibnamefont {MacDonald}},\ }\bibfield  {title} {\bibinfo {title} {Cw three-wave mixing in single-mode optical fibers},\ }\href {https://pubs.aip.org/aip/jap/article/49/10/5098/506705/cw-three-wave-mixing-in-single-mode-optical-fibers} {\bibfield  {journal} {\bibinfo  {journal} {J. Appl. Phys.}\ }\textbf {\bibinfo {volume} {49}},\ \bibinfo {pages} {5098} (\bibinfo {year} {1978})}\BibitemShut {NoStop}%
\bibitem [{\citenamefont {Zorin}(2016)}]{zorin2016josephson}%
  \BibitemOpen
  \bibfield  {author} {\bibinfo {author} {\bibfnamefont {A.}~\bibnamefont {Zorin}},\ }\bibfield  {title} {\bibinfo {title} {Josephson traveling-wave parametric amplifier with three-wave mixing},\ }\href {https://journals.aps.org/prapplied/abstract/10.1103/PhysRevApplied.6.034006} {\bibfield  {journal} {\bibinfo  {journal} {Phys. Rev. Appl.}\ }\textbf {\bibinfo {volume} {6}},\ \bibinfo {pages} {034006} (\bibinfo {year} {2016})}\BibitemShut {NoStop}%
\bibitem [{\citenamefont {Rao}\ \emph {et~al.}(2023)\citenamefont {Rao}, \citenamefont {Yao}, \citenamefont {Wang}, \citenamefont {Zhang}, \citenamefont {Yu},\ and\ \citenamefont {Lu}}]{rao2023unveiling}%
  \BibitemOpen
  \bibfield  {author} {\bibinfo {author} {\bibfnamefont {J.}~\bibnamefont {Rao}}, \bibinfo {author} {\bibfnamefont {B.}~\bibnamefont {Yao}}, \bibinfo {author} {\bibfnamefont {C.}~\bibnamefont {Wang}}, \bibinfo {author} {\bibfnamefont {C.}~\bibnamefont {Zhang}}, \bibinfo {author} {\bibfnamefont {T.}~\bibnamefont {Yu}},\ and\ \bibinfo {author} {\bibfnamefont {W.}~\bibnamefont {Lu}},\ }\bibfield  {title} {\bibinfo {title} {Unveiling a pump-induced magnon mode via its strong interaction with {Walker} modes},\ }\href {https://journals.aps.org/prl/abstract/10.1103/PhysRevLett.130.046705} {\bibfield  {journal} {\bibinfo  {journal} {Phys. Rev. Lett.}\ }\textbf {\bibinfo {volume} {130}},\ \bibinfo {pages} {046705} (\bibinfo {year} {2023})}\BibitemShut {NoStop}%
\bibitem [{\citenamefont {Serha}\ \emph {et~al.}(2025)\citenamefont {Serha}, \citenamefont {McAllister}, \citenamefont {Majcen}, \citenamefont {Knauer}, \citenamefont {Reimann}, \citenamefont {Dubs}, \citenamefont {Melkov}, \citenamefont {Serga}, \citenamefont {Tyberkevych}, \citenamefont {Chumak},\ and\ \citenamefont {Bozhko}}]{rostyslav2025ultra}%
  \BibitemOpen
  \bibfield  {author} {\bibinfo {author} {\bibfnamefont {R.~O.}\ \bibnamefont {Serha}}, \bibinfo {author} {\bibfnamefont {K.~H.}\ \bibnamefont {McAllister}}, \bibinfo {author} {\bibfnamefont {F.}~\bibnamefont {Majcen}}, \bibinfo {author} {\bibfnamefont {S.}~\bibnamefont {Knauer}}, \bibinfo {author} {\bibfnamefont {T.}~\bibnamefont {Reimann}}, \bibinfo {author} {\bibfnamefont {C.}~\bibnamefont {Dubs}}, \bibinfo {author} {\bibfnamefont {G.~A.}\ \bibnamefont {Melkov}}, \bibinfo {author} {\bibfnamefont {A.~A.}\ \bibnamefont {Serga}}, \bibinfo {author} {\bibfnamefont {V.~S.}\ \bibnamefont {Tyberkevych}}, \bibinfo {author} {\bibfnamefont {A.~V.}\ \bibnamefont {Chumak}},\ and\ \bibinfo {author} {\bibfnamefont {D.~A.}\ \bibnamefont {Bozhko}},\ }\bibfield  {title} {\bibinfo {title} {Ultra-long-living magnons in the quantum limit},\ }\href@noop {} {\bibfield  {journal} {\bibinfo  {journal} {arXiv:2505.22773}\ } (\bibinfo {year} {2025})}\BibitemShut {NoStop}%
\bibitem [{\citenamefont {An}\ \emph {et~al.}(2024)\citenamefont {An}, \citenamefont {Xu}, \citenamefont {Mucchietto}, \citenamefont {Kim}, \citenamefont {Moon}, \citenamefont {Hwang},\ and\ \citenamefont {Grundler}}]{an2024emergent}%
  \BibitemOpen
  \bibfield  {author} {\bibinfo {author} {\bibfnamefont {K.}~\bibnamefont {An}}, \bibinfo {author} {\bibfnamefont {M.}~\bibnamefont {Xu}}, \bibinfo {author} {\bibfnamefont {A.}~\bibnamefont {Mucchietto}}, \bibinfo {author} {\bibfnamefont {C.}~\bibnamefont {Kim}}, \bibinfo {author} {\bibfnamefont {K.-W.}\ \bibnamefont {Moon}}, \bibinfo {author} {\bibfnamefont {C.}~\bibnamefont {Hwang}},\ and\ \bibinfo {author} {\bibfnamefont {D.}~\bibnamefont {Grundler}},\ }\bibfield  {title} {\bibinfo {title} {Emergent coherent modes in nonlinear magnonic waveguides detected at ultrahigh frequency resolution},\ }\href {https://www.nature.com/articles/s41467-024-51483-7} {\bibfield  {journal} {\bibinfo  {journal} {Nat. Comms.}\ }\textbf {\bibinfo {volume} {15}},\ \bibinfo {pages} {7302} (\bibinfo {year} {2024})}\BibitemShut {NoStop}%
\bibitem [{\citenamefont {Makiuchi}\ \emph {et~al.}(2021)\citenamefont {Makiuchi}, \citenamefont {Hioki}, \citenamefont {Shimazu}, \citenamefont {Oikawa}, \citenamefont {Yokoi}, \citenamefont {Daimon},\ and\ \citenamefont {Saitoh}}]{makiuchi2021parametron}%
  \BibitemOpen
  \bibfield  {author} {\bibinfo {author} {\bibfnamefont {T.}~\bibnamefont {Makiuchi}}, \bibinfo {author} {\bibfnamefont {T.}~\bibnamefont {Hioki}}, \bibinfo {author} {\bibfnamefont {Y.}~\bibnamefont {Shimazu}}, \bibinfo {author} {\bibfnamefont {Y.}~\bibnamefont {Oikawa}}, \bibinfo {author} {\bibfnamefont {N.}~\bibnamefont {Yokoi}}, \bibinfo {author} {\bibfnamefont {S.}~\bibnamefont {Daimon}},\ and\ \bibinfo {author} {\bibfnamefont {E.}~\bibnamefont {Saitoh}},\ }\bibfield  {title} {\bibinfo {title} {Parametron on magnetic dot: Stable and stochastic operation},\ }\href {https://pubs.aip.org/aip/apl/article/118/2/022402/39849/Parametron-on-magnetic-dot-Stable-and-stochastic} {\bibfield  {journal} {\bibinfo  {journal} {Appl. Phys. Lett.}\ }\textbf {\bibinfo {volume} {118}} (\bibinfo {year} {2021})}\BibitemShut {NoStop}%
\bibitem [{\citenamefont {Elyasi}\ \emph {et~al.}(2022)\citenamefont {Elyasi}, \citenamefont {Saitoh},\ and\ \citenamefont {Bauer}}]{elyasi2022stochasticity}%
  \BibitemOpen
  \bibfield  {author} {\bibinfo {author} {\bibfnamefont {M.}~\bibnamefont {Elyasi}}, \bibinfo {author} {\bibfnamefont {E.}~\bibnamefont {Saitoh}},\ and\ \bibinfo {author} {\bibfnamefont {G.~E.}\ \bibnamefont {Bauer}},\ }\bibfield  {title} {\bibinfo {title} {Stochasticity of the magnon parametron},\ }\href {https://journals.aps.org/prb/abstract/10.1103/PhysRevB.105.054403} {\bibfield  {journal} {\bibinfo  {journal} {Phys. Rev. B}\ }\textbf {\bibinfo {volume} {105}},\ \bibinfo {pages} {054403} (\bibinfo {year} {2022})}\BibitemShut {NoStop}%
\bibitem [{\citenamefont {Gonz{\'a}lez}\ \emph {et~al.}(2024)\citenamefont {Gonz{\'a}lez}, \citenamefont {Litvinenko}, \citenamefont {Kumar}, \citenamefont {Khymyn},\ and\ \citenamefont {{\AA}kerman}}]{gonzalez2024spintronic}%
  \BibitemOpen
  \bibfield  {author} {\bibinfo {author} {\bibfnamefont {V.~H.}\ \bibnamefont {Gonz{\'a}lez}}, \bibinfo {author} {\bibfnamefont {A.}~\bibnamefont {Litvinenko}}, \bibinfo {author} {\bibfnamefont {A.}~\bibnamefont {Kumar}}, \bibinfo {author} {\bibfnamefont {R.}~\bibnamefont {Khymyn}},\ and\ \bibinfo {author} {\bibfnamefont {J.}~\bibnamefont {{\AA}kerman}},\ }\bibfield  {title} {\bibinfo {title} {Spintronic devices as next-generation computation accelerators},\ }\href {https://www.sciencedirect.com/science/article/pii/S1359028624000391} {\bibfield  {journal} {\bibinfo  {journal} {Opin. Solid State Mater. Sci.}\ }\textbf {\bibinfo {volume} {31}},\ \bibinfo {pages} {101173} (\bibinfo {year} {2024})}\BibitemShut {NoStop}%
\end{thebibliography}%

\end{document}

\title{Supplementary Information of “Magnon - Magnon interaction induced by nonlinear spin wave dynamics”}
\author{Matteo Arfini}
\thanks{Equal contribution}
\email{m.arfini@tudelft.nl}
\affiliation{%
Kavli Institute of Nanoscience, Delft University of Technology, Lorentzweg 1,2628 CJ Delft, Netherlands
}
\author{Alvaro Bermejillo-Seco}
\thanks{Equal contribution}
\email{a.bermejilloseco@tudelft.nl}
\affiliation{%
Kavli Institute of Nanoscience, Delft University of Technology, Lorentzweg 1,2628 CJ Delft, Netherlands
}
\author{Artem Bondarenko}
\affiliation{%
Kavli Institute of Nanoscience, Delft University of Technology, Lorentzweg 1,2628 CJ Delft, Netherlands
}
\author{Clinton A. Potts}
\affiliation{Niels Bohr Institute, University of Copenhagen, Blegdamsvej 17, 2100 Copenhagen, Denmark
}
\affiliation{NNF Quantum Computing Programme, Niels Bohr Institute, University of Copenhagen, Denmark}
\author{Yaroslav M. Blanter}
\affiliation{%
Kavli Institute of Nanoscience, Delft University of Technology, Lorentzweg 1,2628 CJ Delft, Netherlands
}
\author{Herre S.J. van der Zant}
\affiliation{%
Kavli Institute of Nanoscience, Delft University of Technology, Lorentzweg 1,2628 CJ Delft, Netherlands
}
\author{Gary A. Steele}
\email{g.steele@tudelft.nl}
\affiliation{%
Kavli Institute of Nanoscience, Delft University of Technology, Lorentzweg 1,2628 CJ Delft, Netherlands
}

\date{\today}

\maketitle

\tableofcontents

\section{Sample and Measurement Setup}
\begin{figure}
    \centering
    \includegraphics[width=0.9\linewidth]{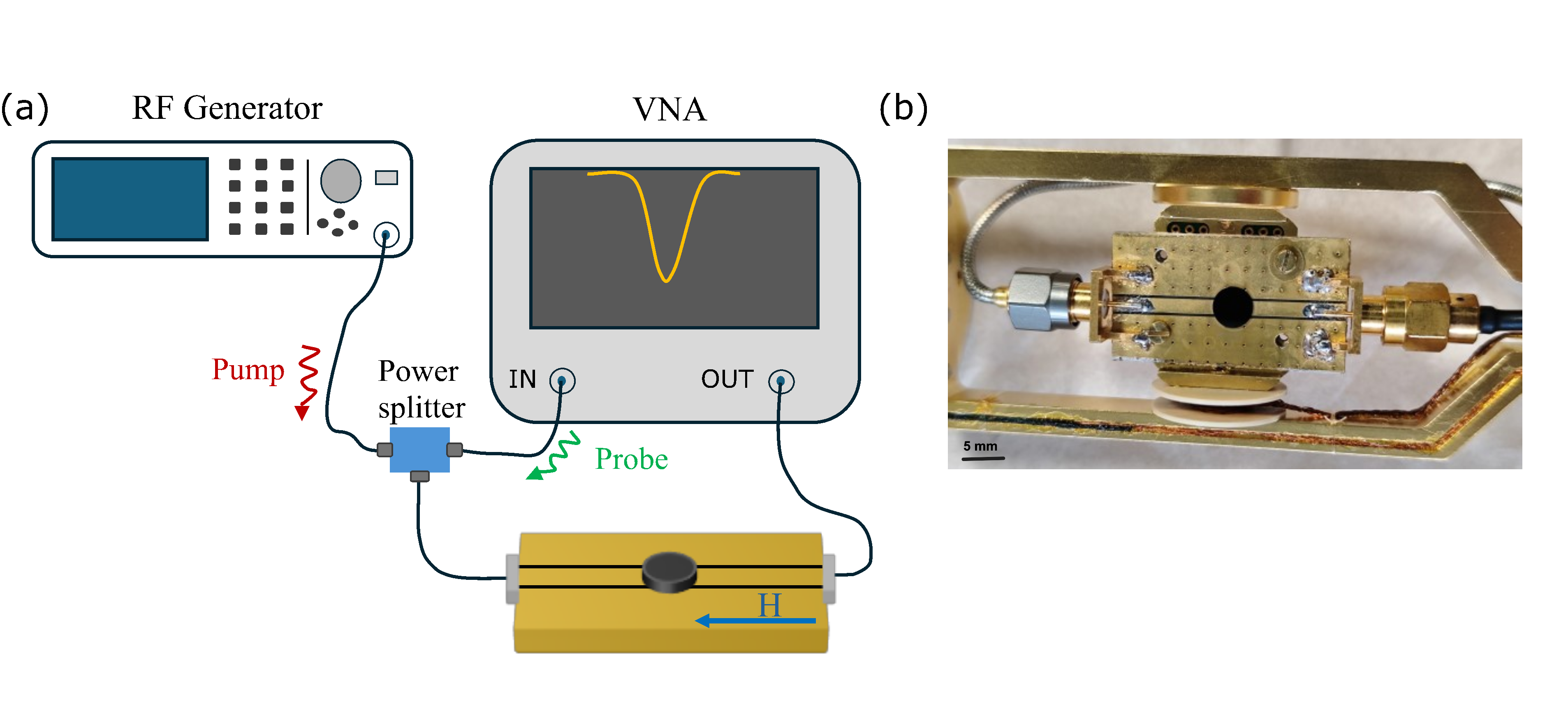}
    \caption{Experimental setup. (a) Schematic of the two-tone measurement configuration. A strong pump drive from an RF generator combines with a weak probe tone from the VNA into the sample. The FMR response of the YIG disk is recorded via a $S_{21}$ measurement. (b) Photograph of the sample mounted on the measurement stick.}
    \label{fig:sample_setup}
\end{figure}

The current experiment is conducted on a commercially available double-side polished, 350 $\mu$m thick, 5 mm yttrium iron garnet (YIG) disk from MTI corporation. The disk is glued with PMMA resist to the center conductor of a 50 $\Omega$ transmission line etched on a standard printed circuit board (PCB), as shown in Fig. \ref{fig:sample_setup}. The sample is then fixed to an RF probe with input and output coaxial cables that enable performing spectroscopy measurements in transmission configuration by probing the $S_{21}$ response from the PCB. The probe is placed in a He environment inside an attoDRY2100 cryostat, equipped with a tunable superconducting magnet up to 9 T. The orientation of the sample is adjusted to ensure that the external bias field is applied along the in-plane orientation of the disk.\\ The FMR resonance spectrum as a function of magnetic field is measured at fixed low input power by performing a vector network analyzer (VNA) sweep for each B field value. The results in the main text are obtained with a two-tone measurement setup shown in Fig. \ref{fig:sample_setup} (a). We use an RF generator to combine in a power splitter (with 3dB attenuation) a strong microwave pump drive with a weak probe tone from the input port of the VNA. The output transmission is then recorded for each value of pump power or frequency.

\section{Ferromagnetic Resonance Spectrum}
\label{sec:fmr}
\begin{figure}
\centering
\includegraphics[width=\linewidth]{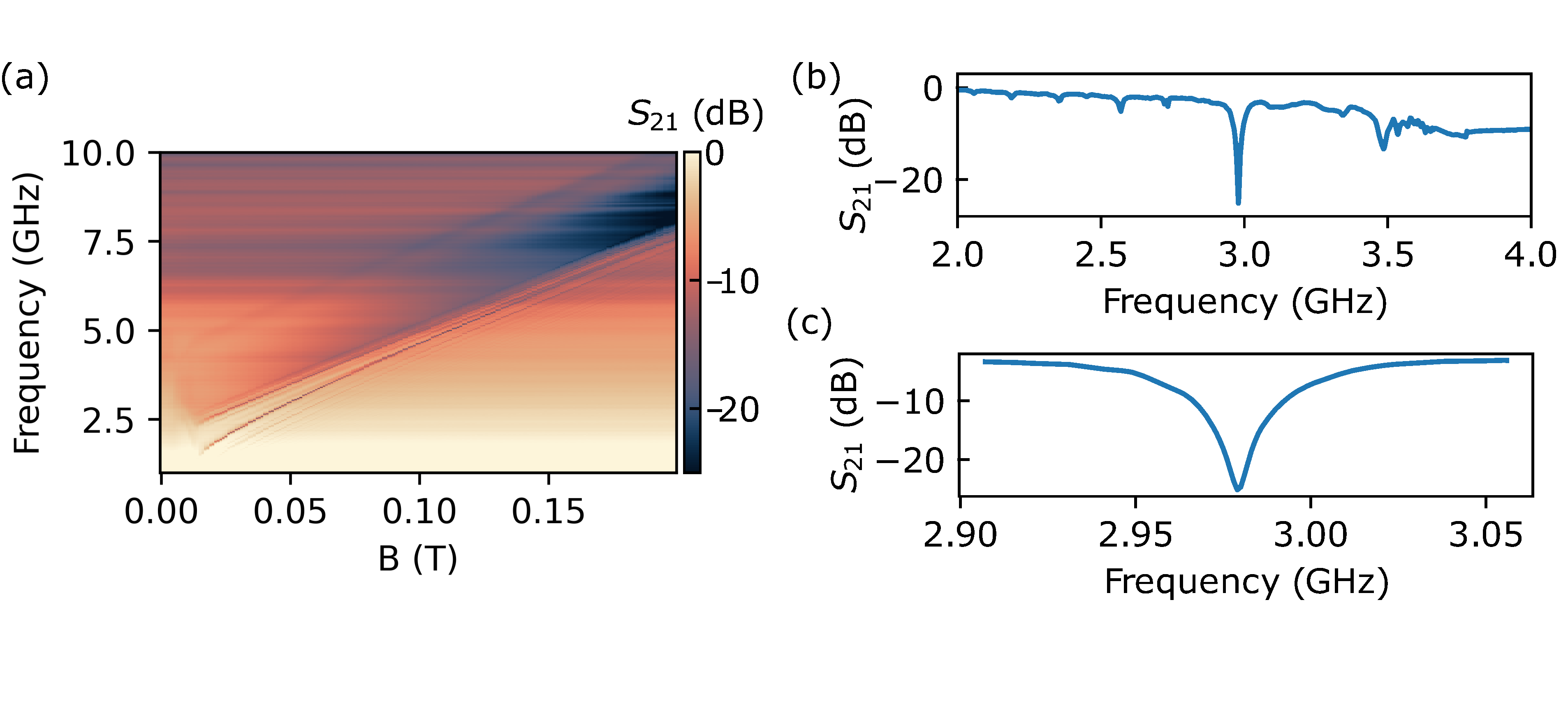}
\caption{Measured FMR spectrum of the YIG disk. (a) Magnetic field dependence of the FMR spectrum from 0 to 200 mT. (b) FMR trace at 50 mT, showing multiple modes. (c) Zoomed-in view of the main resonance at 50 mT.}
\label{fig:FMR}
\end{figure}

The ferromagnetic resonance (FMR) spectrum of the YIG disk is presented in Fig.\ref{fig:FMR}. All observed modes exhibit a magnetic field dependence consistent with Kittel's formula: $\omega=\gamma\mu_0\sqrt{(H(H+M_S)}$. To find the exact dependence, one would need to find a way to account for the demagnetization field of the sample. Fig.~\ref{fig:FMR}(b) shows a representative spectrum at 50 mT, where multiple FMR modes are visible, corresponding to the Walker modes of the disk. Among these, the mode with the highest intensity, which couples most strongly to the feedline, is highlighted in Fig.~\ref{fig:FMR}(c). This is the mode analyzed in detail in the main text.

To better understand the Walker modes in the in-plane magnetized disk, we performed numerical simulations using Mumax3 \cite{Vansteenkiste2014,Exl2014}. The simulations account for dipole-dipole interactions by computing the demagnetization field and solving the spin dynamics within this field. Exchange interactions were deliberately omitted, since they are known to not play a role at the relevant wavelengths. The disk-shaped particle was discretized in 128x128x32 grid, with the cell size of 41$\mu$mx41$\mu$mx11$\mu$m. An external magnetic field of 50 mT was applied along the y-direction, and an initial magnetization aligned with the same direction.

\begin{figure}
\centering
\includegraphics[width=0.8\linewidth]{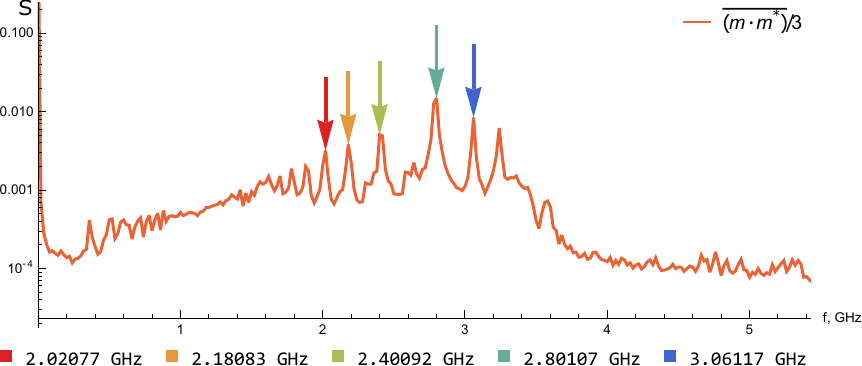}
\caption{Simulated FMR spectrum of the YIG disk at 50 mT, obtained using Mumax3. Colored arrows indicate identified modes. The highest intensity resonance, at 2.8 GHz, corresponds to the mode analyzed in the main text.}
\label{fig:mumax-trace}
\end{figure}

The system was excited by an approximated field of a strip antenna, with the 50 mT DC bias applied along the antenna direction, like in the experiment. We use a wideband sinc pulse with a limit frequency of $15$ GHz to limit aliasing just to the upper portion of the spectrum far from the modes of interest. Mumax3 computes the time evolution of the magnetization, and the resulting spectra were obtained via Fast Fourier Transforms (FFTs). Figure~\ref{fig:mumax-trace} shows the simulated spectrum, where we plot a volume average of magnetic precession intensity to catch all the excited modes. Several Walker modes are resolved. The dominant mode appears at 2.8 GHz, which closely matches the experimentally observed frequency of 2.98 GHz. The slight discrepancy likely arises from the coarse spatial discretization imposed by computational limitations.

\begin{figure}
    \centering
    \includegraphics[width=0.8\linewidth]{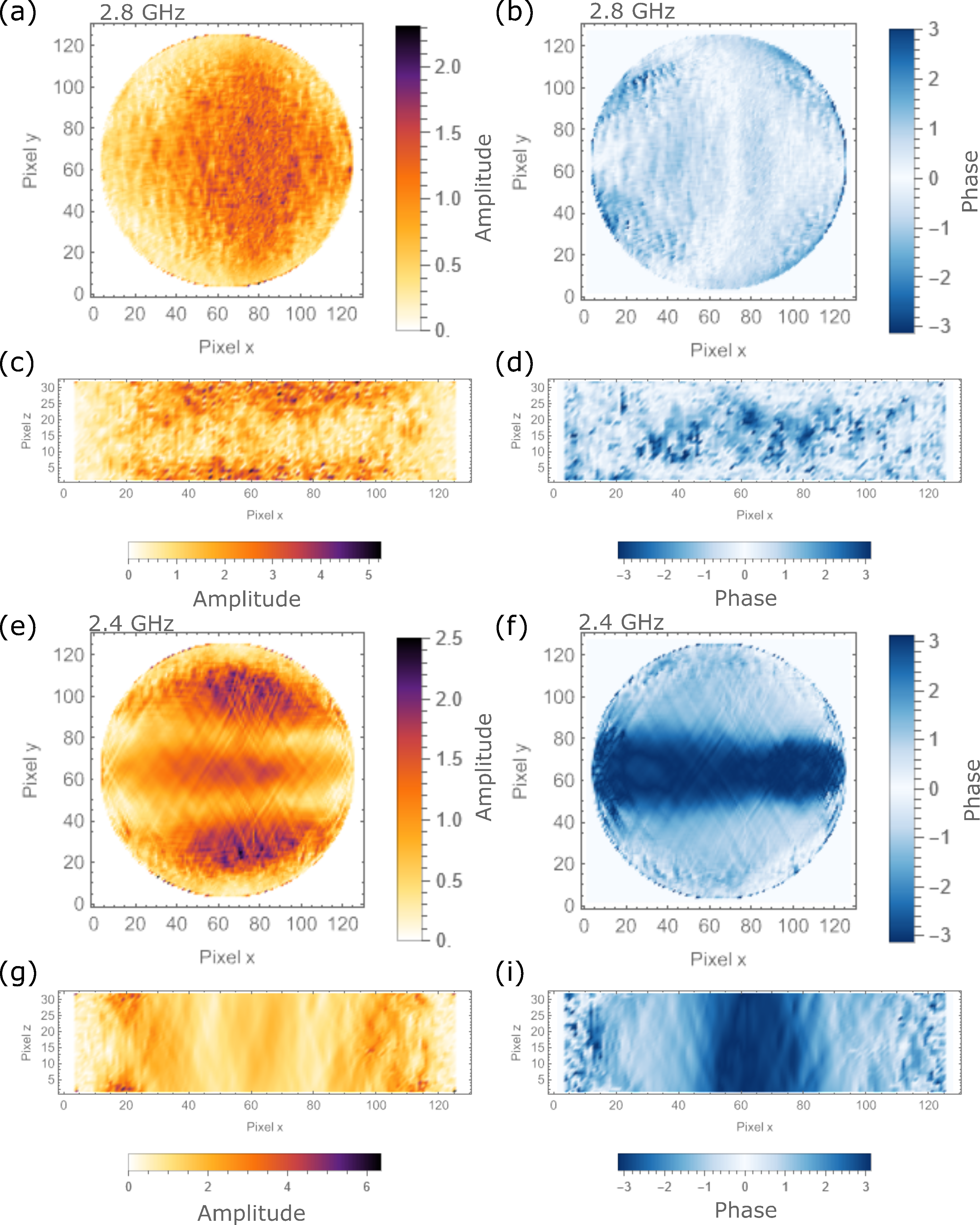}
    \caption{Mode profiles of a YIG disk computed with Mumax3. (a) and (e) are maps of the absolute intensity in a slice of the disk for mode frequencies 2.4 and 2.8 GHz, respectively. (c) and (g) are analogous, with slices through the thickness of the disk. (b), (d), (f), and (i) are maps of the phases, in rad, corresponding to the intensity map on their left. The pixels correspond to the 128×128×32 simulation space with lengths per pixel 41$\mu$mx41$\mu$mx11$\mu$m.}
    \label{fig:mumax-color}
\end{figure}

To characterize the spatial profiles of the modes, we excited the system with pure sine waves at the frequencies of the identified modes. The corresponding mode profiles are shown in Fig.\ref{fig:mumax-color}. The dominant mode at 2.8 GHz exhibits a uniform profile both along the disk's radius and through its thickness. Lower frequency modes, such as the one at 2.4 GHz, display a higher number of nodes along the y-direction while remaining homogeneous across the thickness. These features are evident in both the amplitude and phase maps of the modes, as depicted in Fig.\ref{fig:mumax-color}(a)-(i).

\section{Theory}
\label{sec:theory}

As discussed in the main text, the Suhl instability results in a three-wave scattering process between the strongly driven magnetostatic mode and a pair of propagating modes at half the frequency. In this section, we will outline the derivation of the dispersion relationship of the propagating modes, describe the theoretical model, and extract the three-wave mixing coupling rate. 

\subsection{Dispersion Relationship and coupling rate}
\label{sec:dispersion}
Understanding the dispersion relation in the sample is crucial to identifying the magnon ($\hat{m}_k$) modes available for three-wave mixing that satisfy energy and momentum conservation. Analytical expressions for the dispersion relation of an in-plane magnetized thin disk have not been comprehensively addressed in the literature. M. Sparks \cite{Sparks1970} described dispersion relations for out-of-plane magnetized disks and some specific modes in in-plane magnetized configurations. The primary challenge lies in solving the Walker equations to accurately determine the sample’s demagnetization field. However, given the large radius and thickness of our sample, the dispersion relation can be approximated by that of an ellipsoid as a reasonable first-order approximation. Using a frame of reference with $z$ in the out of plane direction and the feedline and magnetic field aligned along $x$, we can define \cite{l2012waveturbulence}
\begin{equation}
    A_k = \mu_0 \gamma \left(H + M_S \lambda_{ex} k^2 + M_S \frac{k_y^2 + k_z^2}{2k^2}\right), \quad B_k = \mu_0 \gamma M_S \frac{(k_y + i k_z)^2}{2k^2},
\end{equation}
such that
\begin{equation}
    \omega_k=\sqrt{A_k^2-|B_k|^2}.
    \label{eq:dispersion}
\end{equation}
Here, $\mu_0$ is the permeability of free space, $\gamma$ is the gyromagnetic ratio, $H$ is the external magnetic field, $M_S$ is the saturation magnetization, and $\lambda_{\text{ex}}$ is the exchange stiffness constant and k the modulus of the momentum ($k=\sqrt{k_x^2+k_y^2+k_z^2}$).

\begin{figure}
    \centering
    \includegraphics[width=0.8\linewidth]{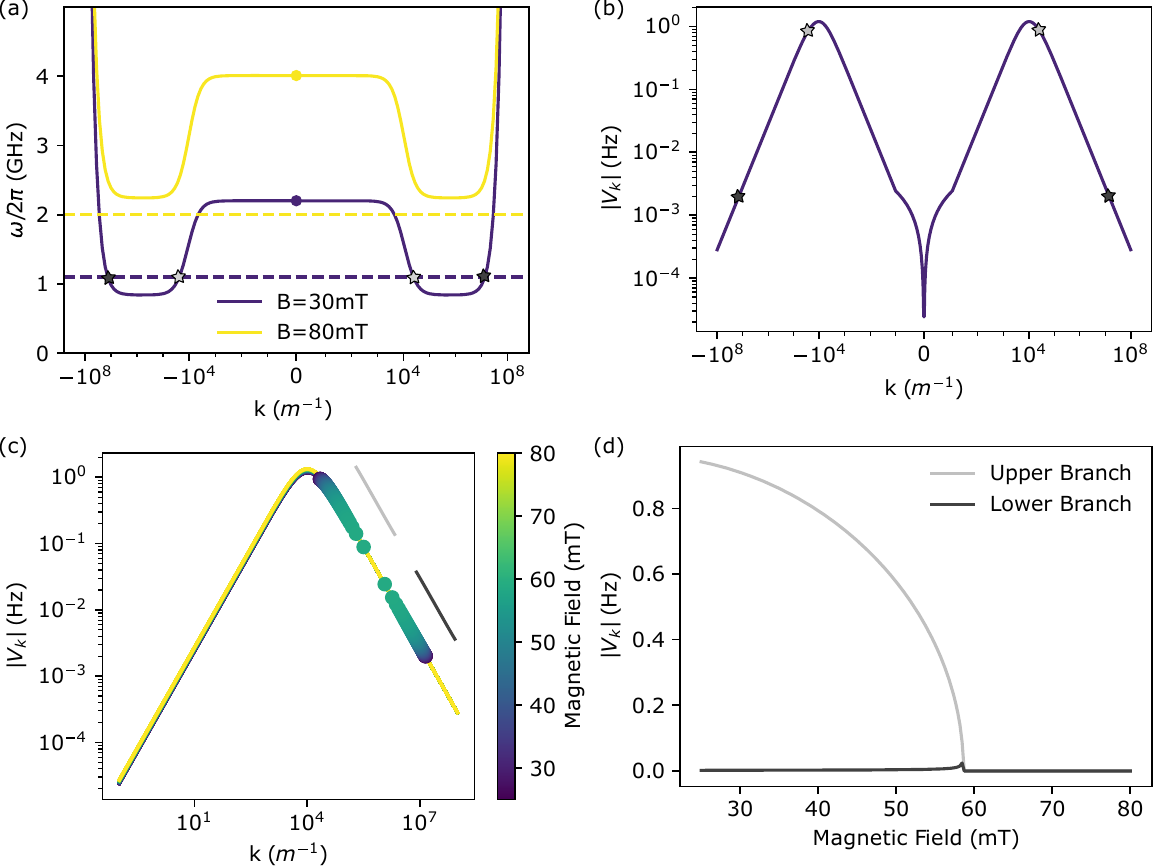}
    \caption{Calculation of the three-wave mixing coupling. (a) Dispersion relation of the backward volume mode with $k_z=\pi/t$ at fields below the threshold (30 mT) and above the threshold (80 mT), calculated from Eq. (\ref{eq:dispersion}). (b) Calculation of the coupling as a function of in-plane momentum k (=$k_x$) at 30 mT. The allowed k values from (a) are also indicated here with the same star symbols. (c) Magnetic field dependence of the allowed k modes. The curve $V_k$ does not vary significantly, but the values of k at which the energy conservation applies do, as shown by the dots, color-coded to show the magnetic field. (d) Magnetic field dependence of the coupling for the upper and lower branches, indicated with colored lines in (c). The upper branch, corresponding to the lower k modes, has significantly higher coupling up to the magnetic field threshold.}
    \label{fig:SIcoupling}
\end{figure}

To have available states for three-wave mixing, it is required to have a dispersion that has lower frequencies for increasing k in some range of the dispersion. That happens for any branch other than the Damon-Eschbach branch ($k_x=0$), and is most pronounced for $k_y=0$, that is, for backward volume magnetostatic waves. An example of the dispersion at $k_y=k_z=0$, for two different values of the external magnetic field, is shown in Figure 4 of the main text, and repeated in Fig.~ \ref{fig:SIcoupling} for clarity. Here we can see that only at fields below the critical field are available modes at half the frequency of $\omega(k=0)$. 

The coupling of a magnetostatic mode to a pair of such k states was calculated for a sufficiently large ellipsoidal sample in \cite{l2012waveturbulence}, and is given by
\begin{equation}
    V_k = \omega_M \sqrt{\frac{g_{eff} \mu_B}{2 \nu M_S}} \left(1 + \frac{\omega_k}{A_k + |B_k|}\right) \frac{k_x (k_y + i k_z)}{k^2}, \quad \omega_M = \mu_0 \gamma M_S.
\end{equation}
The first insight this equation provides is that for pure backward volume spin waves, i.e., $k$ is parallel to the magnetisation, $k_y=k_z=0$, the coupling is zero. The situation for which one finds the higher coupling is for $k_y=0$, and $k_z=\pi/t$, corresponding to one node along the thickness direction. This result does not provide accurate values of the coupling rate in our geometry, due to the mismatch in the considered demagnetisation field. Still, it enables us to get an order-of-magnitude estimate of the coupling and evaluate which of the two pairs of magnons that are enabled by E-k conservation participate more strongly in the process. These results are gathered in Fig.~\ref{fig:SIcoupling}. Fig.~\ref{fig:SIcoupling}(a) shows how we determine which values of k are allowed by E-k conservation at a field of $H=$30 mT. Fig.~\ref{fig:SIcoupling}(b) shows the k-dependent coupling rate, where the allowed k-modes are highlighted with stars. We observe that the lower k-mode (in bright gray) has a rate 3 orders of magnitude larger than the higher k-mode (in dark gray). Thus, we can safely discard the latter when doing our analysis. Finally, in Fig.~\ref{fig:SIcoupling}(d) we show the evolution of the coupling with the B field as shown in Fig. 4 (b) of the main text. 

These calculations only provide order-of-magnitude estimates of the coupling and understanding of the participating modes. To get a better estimate, even within the ellipsoidal sample approximation, one would need to consider all k-modes for which the E-k condition is satisfied. Those will be a continuum along $k_y$ and a sum of discrete modes with $k_z=n\pi/t$. 

\subsection{Three-wave mixing Hamiltonian}
\label{sec:Hamiltonian}
 In the main text, we stated that the effective Hamiltonian results in a beam splitter interaction and squeezing of the $\pm k$ modes. Here we show all the necessary steps to arrive at that result. To begin, we will consider a quasi-magnetostatic mode $\hat{m}_0$, which is coupled via a three-wave mixing process to pairs of propagating modes $\hat{m}_{\pm k}$ living within the spin wave continuum with momentum $\vert k\vert \neq 0$. The coupling was originally proposed by Suhl and is described by the Hamiltonian \cite{suhl1957theory}, 
\begin{equation}
    \hat{\mathcal{H}}/\hbar = \omega_{0}\hat{m}_0^{\dagger}\hat{m}_0 + \sum_{k>0} \big( \omega_{\rm k} \hat{m}_{k}^{\dagger}\hat{m}_{k} + \omega_{\rm -k} \hat{m}_{-k}^{\dagger}\hat{m}_{-k} + V_k^*\hat{m}_0\hat{m}_{k}^{\dagger}\hat{m}_{-k}^{\dagger} +V_k\hat{m}_0^{\dagger}\hat{m}_{k}\hat{m}_{-k} \big) + \hat{\mathcal{H}}_{\rm d}/\hbar.
    \label{eq:origHam}
\end{equation}
Here $g(\vert k \vert)$ is the coupling rate between the magnetostatic mode and the continuum modes of wavenumber $\vert k \vert$. Moreover, we consider a direct coherent drive of the magnetostatic mode of the form,
\begin{equation}
    \hat{\mathcal{H}}_{\rm d}/\hbar = i(\Omega_d^*\hat{m}_0 e^{i\omega_d t} -\Omega_d \hat{m}_0^\dagger e^{-i\omega_d t}),
\end{equation}
where $\Omega_d$ is the drive strength, and $\omega_d$ is the drive frequency. The time dependence can be removed from the drive by moving to a frame rotating with the drive via the unitary $U(t)=e^{i\omega_d \hat{m}_0^\dagger \hat{m}_0 t}$, giving a new Hamiltonian of the form,
\begin{equation}
    \hat{\mathcal{H}}/\hbar =\Delta_{0}\hat{m}_0^{\dagger}\hat{m}_0 + \sum_{k>0} \big( \omega_{\rm k} \hat{m}_{k}^{\dagger}\hat{m}_{k} + \omega_{\rm -k} \hat{m}_{-k}^{\dagger}\hat{m}_{-k} + V_k^*\hat{m}_0e^{-i\omega_d t}\hat{m}_{k}^{\dagger}\hat{m}_{-k}^{\dagger} + V_k\hat{m}_0^{\dagger}e^{i\omega_d t}\hat{m}_{k}\hat{m}_{-k} \big) + i(\Omega_d^*\hat{m}_0 -\Omega_d \hat{m}_0^\dagger).
\end{equation}
Here we define the detuning as $\Delta_0 =  \omega_0 - \omega_d$. Following the above discussion, from now on, we retain the coupling to a single pair of modes at $k$ and $-k$. To remove the time-dependence from the Hamiltonian, we can move to a frame co-rotating with the frequency of the continuum mode via $U(t)=e^{i\omega_d/2 \hat{m}_k^\dagger \hat{m}_k t}e^{i\omega_d/2 \hat{m}_{-k}^\dagger \hat{m}_{-k} t}$ resulting in the Hamiltonian of the form
\begin{equation}
    \hat{\mathcal{H}}/\hbar =\Delta_{0}\hat{m}_0^{\dagger}\hat{m}_0 + \Delta_{\rm k} \hat{m}_{k}^{\dagger}\hat{m}_{k} + \Delta_{\rm -k} \hat{m}_{-k}^{\dagger}\hat{m}_{-k} + V_k^*\hat{m}_0\hat{m}_{k}^{\dagger}\hat{m}_{-k}^{\dagger} + V_k\hat{m}_0^{\dagger}\hat{m}_{k}\hat{m}_{-k} + i(\Omega_d^*\hat{m}_0 -\Omega_d \hat{m}_0^\dagger).
    \label{eq:rotHam}
\end{equation}
Here we have defined $\Delta_{\pm k} = \omega_{\pm k} - \omega_0/2$.

\subsection{Steady-State Solutions}
\label{sec:steadystate}
Using the Hamiltonian defined in Eq.~\ref{eq:rotHam} we derived the dynamics of an arbitrary operator $\hat{\mathcal{O}}$ via the Heisenberg equation $-i\hbar \frac{d\hat{\mathcal{O}}}{dt} =[\hat{\mathcal{H}},\hat{\mathcal{O}}]$ with the addition of dissipation terms due to interactions with the environment. Since we operate in the high temperature limit where $k_BT\gg \hbar\omega$, we can ignore quantum fluctuations, and obtain the following semi-classical equations of motion:
\begin{equation}
    \langle \dot{\hat{m}}_0 \rangle = -(i\Delta_0 + \gamma_{0}/2)\langle \hat{m}_0 \rangle - iV_k \langle \hat{m}_k \rangle \langle \hat{m}_{-k} \rangle - \Omega_d,
\end{equation}
\begin{equation}
    \langle \dot{\hat{m}}_{\pm k} \rangle = -(i\Delta_{\pm k} + \gamma_{\pm k}/2)\langle \hat{m}_{\pm k} \rangle -iV_k^* \langle \hat{m}_0\rangle \langle \hat{m}_{\mp k}^\dagger\rangle.
\end{equation}
Here $\gamma_0$ is the decay rate of the magnetostatic mode and $\gamma_{\pm k}$ is the decay of the continuum mode, which we have assumed to be equal for the forward and backward propagating modes. Moreover, we have performed a mean-field approximation where we have assumed $\langle \hat{\mathcal{A}}\hat{\mathcal{B}}\rangle = \langle \hat{\mathcal{A}} \rangle \langle \hat{\mathcal{B}} \rangle$. The classical steady-state values can be obtained by setting the time derivatives equal to zero, resulting in:
\begin{equation}
    \langle \hat{m}_0 \rangle = -\frac{\Omega_d + iV_k\langle \hat{m}_k\rangle\langle \hat{m}_{-k}\rangle}{i\Delta_0 + \gamma_{0}/2}, \qquad
    \langle \hat{m}_{k} \rangle = -\frac{iV_k^*\langle \hat{m}_0\rangle \langle \hat{m}_{- k}^\dagger\rangle}{i\Delta_{k} + \gamma_{k}/2}, \qquad
    \langle \hat{m}_{-k} \rangle = -\frac{iV_k^*\langle \hat{m}_0\rangle \langle \hat{m}_{k}^\dagger\rangle}{i\Delta_{-k} + \gamma_{-k}/2}.
    \label{eq:steady}
\end{equation}
Due to the symmetry of the scattering, we will consider the case where the classical steady-state values of the continuum mode are identical: $\vert\langle \hat{m}_{k} \rangle\vert = \vert\langle \hat{m}_{-k} \rangle\vert = \beta$. Moreover, if we consider the steady-state values to have the form $\langle\hat{m}_k\rangle = \beta e^{i\phi_k}$, $\langle\hat{m}_{-k}\rangle = \beta e^{i\phi_{-k}}$, and we write the coupling in the form $V_k = \vert V_k \vert e^{i\phi_g}$, we can define a phase $\phi = \phi_k + \phi_{-k} + \phi_g$, we have chosen a gauge where $\langle\hat{m}_0\rangle$ is real. Therefore, we can rewrite the steady state equations in the form,
\begin{equation}
\begin{split}
    &(i\Delta_0 + \gamma_{0}/2) \langle \hat{m}_0 \rangle + \Omega_d + i \vert V_k \vert \beta^2 e^{i\phi}= 0 \\
    &(i\Delta_k + \gamma_k/2 ) \beta e^{i\phi} + i\vert V_k \vert \langle \hat{m}_0 \rangle \beta = 0
\end{split}
\end{equation}
These equations can be solved for $\beta$,
\begin{equation}
        (i\Delta_k + \gamma_k/2)({i\Delta_0 + \gamma_{0}/2}) \beta e^{i\phi}  =ig \Omega_d \beta- V_k^2 \vert \beta \vert^2 \beta e^{i\phi}.
\end{equation}
In the relevant situation of zero detuning $\Delta_0 = \Delta_k$, we can see that $\beta$ has real solutions if $\phi = \pi/2$. Under these assumptions, the steady state value of the continuum mode can be solved for explicitly, giving
\begin{equation}
    \beta = \sqrt{\frac{4V_k\Omega_d-\gamma_{0}\gamma_k}{4V_k^2}}
    \label{beta_amplitude}
\end{equation}
We can immediately notice that the continuum mode requires a non-zero amplitude for drives above the critical amplitude
\begin{equation}
    \Omega_d > \frac{\gamma_k\gamma_0}{4V_k}.
\end{equation}
This power corresponds to the onset of the parametric generation of modes within the continuum. At the threshold power where $\beta =0$, we can solve for the steady state value of the Kittle mode 
\begin{equation}
    \langle \hat{m}_0\rangle_{cr} = \frac{\Omega_d}{\gamma_{0}/2} = \frac{\gamma_k}{2V_k},
\end{equation}
which is known as the critical amplitude of the $k=0$ magnetostatic mode. 

\subsection{Linearized Dynamics}
\label{sec:lineardynamics}

We can now consider the dynamics of the fluctuations about the steady-state values: $\hat{\mathcal{O}} = \langle\hat{\mathcal{O}}\rangle + \delta\hat{\mathcal{O}}$. Neglecting higher-order terms in the fluctuations, we can rewrite the Hamiltonian. If we consider the term $\hat{m}_0\hat{m}_k^{\dagger}\hat{m}_{-k}^{\dagger}$, we assume here that $V_k$ and $\langle\hat{m}_k\rangle = \beta$ are real. As discussed in the section above, in the gauge of choice we took $\phi=\phi_{-k}=\pi/2$ which leads to $\langle\hat{m}_{-k}\rangle = i\beta$. We can expand about these steady-state values,
\begin{equation}
\begin{split}
 \hat{m}_0\hat{m}_k^\dagger \hat{m}_{-k}^\dagger\rightarrow (\langle \hat{m}_0\rangle+\delta \hat{m}_{0})(\beta+\delta \hat{m}^\dagger_{k})(-i\beta+\delta \hat{m}^\dagger_{-k})=\\
 (\langle \hat{m}_0 \rangle \delta \hat{m}^\dagger_{k}+\delta \hat{m}_{0}\beta+\delta \hat{m}_{0}\delta \hat{m}^\dagger_{k}+\langle \hat{m}_0\rangle\beta )(-i\beta+\delta \hat{m}^\dagger_{-k})=\\
 -i\langle \hat{m}_0 \rangle \delta \hat{m}^\dagger_{k}\beta+\langle \hat{m}_0 \rangle \delta \hat{m}^\dagger_{k}\delta \hat{m}^\dagger_{-k}-i\delta \hat{m}_{0} \vert \beta \vert^2 +\beta\delta \hat{m}_{0}\delta \hat{m}^\dagger_{-k} \\
 -i\beta\delta \hat{m}_{0}\delta \hat{m}^\dagger_{k}+
 \delta \hat{m}_{0}\delta \hat{m}^\dagger_{k}\delta \hat{m}^\dagger_{-k}-i\langle \hat{m}_0\rangle\vert \beta \vert^2+\langle \hat{m}_0\rangle\beta\delta \hat{m}^\dagger_{-k}.
\end{split}
 \label{lineariz_1}
\end{equation}
Keeping only terms quadratic in fluctuations gives
\begin{equation}
\begin{split}
 \hat{m}_0\hat{m}_k^\dagger \hat{m}_{-k}^\dagger \rightarrow \langle \hat{m}_0 \rangle \delta \hat{m}^\dagger_{k}\delta \hat{m}^\dagger_{-k}+\beta \delta \hat{m}_{0}\delta \hat{m}^\dagger_{-k} -i 
 \beta \delta \hat{m}_{0}\delta \hat{m}^\dagger_{k}.
\end{split}
 \label{lineariz_2}
\end{equation}
The Hamiltonian can be simplified further by performing a Bogoliubov transformation to a collective mode basis defined by:
\begin{equation}
    \hat{m}_{k+} = \frac{1}{\sqrt{2}}(i\delta\hat{m}_k + \delta\hat{m}_{-k}), \qquad
    \hat{m}_{k-} = \frac{1}{\sqrt{2}}(\delta\hat{m}_k + i\delta\hat{m}_{-k}).
\end{equation}
Therefore, the original modes can be redefined in the form
\begin{equation}
    \delta\hat{m}_k = \frac{1}{\sqrt{2}}(-i\hat{m}_{k+}+\hat{m}_{k-}), \qquad
    \delta\hat{m}_{-k} = \frac{1}{\sqrt{2}}(\hat{m}_{k+}-i\hat{m}_{k-}).
\end{equation}
Using these new modes, the interaction terms can be written in the form
\begin{equation}
    \hat{m}_0\hat{m}_k^\dagger \hat{m}_{-k}^\dagger \rightarrow \frac{i\langle \hat{m}_0\rangle}{2}\left((\hat{m}_{k+}^\dagger)^2+(\hat{m}_{k-}^\dagger)^2\right)+\sqrt{2}\beta\delta \hat{m}_0 \hat{m}_{k+}^\dagger,
\end{equation}
from which we automatically find also its Hermitian conjugate
\begin{equation}
    \hat{m}_0^\dagger \hat{m}_k \hat{m}_{-k}\rightarrow \frac{-i\langle \hat{m}_0\rangle}{2}\left((\hat{m}_{k+})^2+(\hat{m}_{k-})^2\right)+\sqrt{2}\beta\delta \hat{m}^\dagger_0 \hat{m}_{k+}.
\end{equation}
Therefore, in the new collective mode basis, we can write the linearized Hamiltonian in the form, 
\begin{equation}
    \begin{aligned}
    \hat{\mathcal{H}}/\hbar = &\Delta \delta \hat{m}^\dagger_0\delta \hat{m}_0+\Delta_+ \hat{m}_{k+}^\dagger \hat{m}_{k+}+\Delta_- \hat{m}_{k-}^\dagger \hat{m}_{k-} \\&- \frac{iV_k}{2}\langle m_0\rangle \left[(\hat{m}_{k+})^2 - (\hat{m}_{k+}^\dagger)^2 +(\hat{m}_{k-})^2 -(\hat{m}_{k-}^\dagger)^2)\right]  + \sqrt{2}V_k\beta \left(\delta \hat{m}_0 \hat{m}_{k+}^\dagger+\delta \hat{m}^\dagger_0 \hat{m}_{k+}\right).
    \end{aligned}
\end{equation}
Where we recognize an effective beam-splitter interaction
\begin{equation}
    \hat{\mathcal{H}}_{int}/\hbar = \sqrt{2}V_k\beta \left(\delta \hat{m}_0 \hat{m}_{k+}^\dagger+\delta \hat{m}^\dagger_0 \hat{m}_{k+}\right)
\end{equation}
that scales with the amplitude of the parametrically pumped collective mode. There exists an additional single-mode squeezing interaction. However, the magnitude of this interaction is fixed due to the critical amplitude of the magnetostatic mode. 

\subsection{Output Spectrum}
\label{sec:spectrum}

If we consider the linearized Hamiltonian, we can derive the quantum Langevin equations, omitting the dark mode $\hat{m}_{k-}$:
\begin{equation}
    \delta\dot{\hat{m}}_0 = -(i\Delta_0 + \gamma_0/2)\delta\hat{m}_0 -i\sqrt{2}V_k\beta\hat{m}_{k+} + \sqrt{\gamma_{\rm ext}}\hat{m}_{\rm in}(t)+\sqrt{\gamma_{\rm int}}\delta\hat{\eta}(t),
    \label{eq:eomdelb}
\end{equation}
\begin{equation}
    \dot{\hat{m}}_{k+} = -(i\Delta_d + \gamma_n/2)\hat{m}_{k+} +\frac{V_k}{2}\langle\hat{m}_0\rangle\hat{m}_{k+}^{\dagger}-i\sqrt{2}V_k\beta\delta\hat{m}_0 + \sqrt{\gamma_k}\hat{\zeta}(t).
    \label{eq:eomdeln}
\end{equation}
These equations describe the dynamics of the fluctuations, and we have explicitly included coupling with the environment via the noise operators $\hat{\eta}(t)$ and $\hat{\zeta}(t)$. Moreover, we have included an external probe tone of drive strength $\varepsilon_p$. Following the standard Markovian approximation, these noise operators are described by the correlators:
\begin{equation}
    \langle \hat{\mathcal{O}}(t)\hat{\mathcal{O}}^{\dagger}(t^{\prime})\rangle = (n_{th} + 1)\delta(t-t^{\prime}),\qquad 
    \langle \hat{\mathcal{O}}^{\dagger}(t)\hat{\mathcal{O}}(t^{\prime})\rangle = n_{th}\delta(t-t^{\prime}),
\end{equation}
where $n_{th} = 1/({\rm exp}(\hbar\omega/k_BT)-1)$ is the average thermal population of a bosonic mode of frequency $\omega$ and temperature $T$. In the experimental configuration $\langle\hat{m}_0\rangle , \beta \gg n_{th}$. Therefore, we will ignore the fluctuations and consider the semi-classical description of the dynamics. 

We can write Eqs.~\ref{eq:eomdelb},\ref{eq:eomdeln} in the frequency domain by performing the Forier transform $\hat{\mathcal{O}}(\omega) = \int_{-\infty}^{\infty}dt e^{i\omega t}\hat{\mathcal{O}}(t)$:
\begin{equation}
    \chi_m^{-1}(\omega) \delta\hat{m}_0(\omega) = -i\sqrt{2}V_k\beta\hat{m}_{k+}(\omega) + \sqrt{\gamma_{\rm ext}}\hat{m}_{\rm in}(\omega),
\end{equation}
\begin{equation}
    \chi_{m+}^{-1}(\omega)\hat{m}_{k+}(\omega) = +\frac{V_k}{2}\langle\hat{m}_0\rangle\hat{m}_{k+}^{\dagger}(-\omega)-i\sqrt{2}V_k\beta\delta\hat{m}_0(\omega).
\end{equation}
Here $\chi_m^{-1} = 1/(i(\Delta_0-\omega)+\gamma_0/2)$ and  $\chi_d^{-1} = 1/(i(\Delta_+-\omega)+\gamma_n/2)$ are the susceptibilities. Using these equations, we can construct a system of equations defined by $XA=B$, with:
\begin{equation}  
A  =
\begin{pmatrix}
    \chi_m^{-1}(\omega) & 0 & i\sqrt{2}V_k\beta & 0 \\
    0 & [\chi_m^{-1}]^*(-\omega) & 0 & -i\sqrt{2}V_k\beta \\
    i\sqrt{2}V_k \beta & 0 &  \chi_{m+}^{-1}(\omega) & V_k\langle m_0\rangle/2 \\
    0 & -i\sqrt{2}V_k \beta & V_k\langle m_0\rangle/2 &  [\chi_{m+}^{-1}]^*(-\omega)
\end{pmatrix}, \qquad
X = \begin{bmatrix} \delta\hat{m}_0(\omega) \\ \delta\hat{m}_0^{\dagger}(-\omega) \\ \hat{m}_{k+}(\omega) \\ \hat{m}_{k+}^{\dagger}(-\omega) \\ \end{bmatrix}, \qquad B = \begin{bmatrix} \sqrt{\gamma_{\rm ext}}\hat{m}_{\rm in} \\ \sqrt{\gamma_{\rm ext}}\hat{m}_{\rm in}^* \\ 0 \\ 0\end{bmatrix}.
\end{equation}

\begin{figure}
    \centering
    \includegraphics[width=\linewidth]{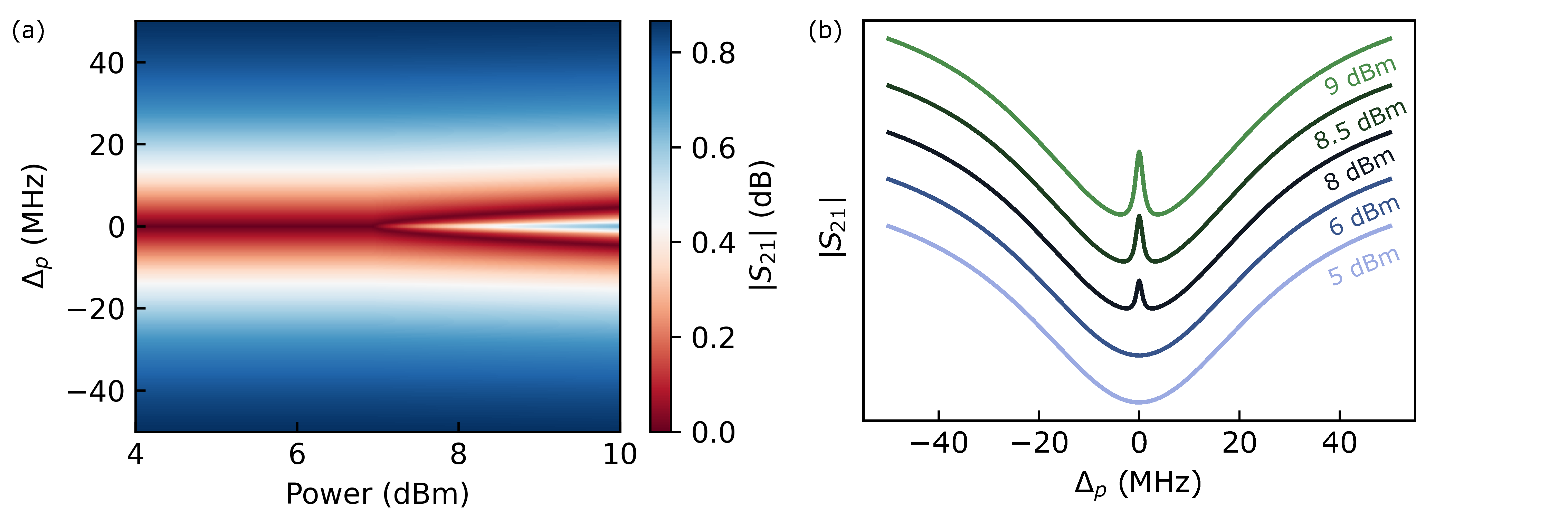}
    \caption{Analytical calculation of the resulting power-dependent FMR spectrum of the driven $\hat{m}_0$ mode in a two-tone experiment. (a) Colormap of the resulting dynamics of the fluctuations around the steady state of the $k=0$ mode when probed by a near-resonant weak drive tone from Eq. \ref{S21_in_out_eq}. The analytical model well captures the appearance of the splitting feature at high power in the probed spectrum. Linecuts of the computed transmission $S_{21}$ are displayed in (b).}
    \label{fig:S21_theory}
\end{figure}

The vector $X$ is solved numerically, and using the input-output formalism, we can calculate the spectral response of the weak probe tone
\begin{equation}
    S_{21}(\omega) = 1 - \sqrt{\gamma_{\rm ext}} \delta\hat{m}_0(\omega)/\hat{m}_{\rm in}(\omega).
    \label{S21_in_out_eq}
\end{equation}
The calculated output spectrum of the $k=0$ mode as a function of input microwave power is shown in Fig. \ref{fig:S21_theory}, for parameters close to the experimental data in the main text. The theoretical predictions exhibit excellent agreement with the measurement data shown in Fig 2 in the main text, revealing the appearance of the spectral splitting feature above the critical power threshold. 
\section{Data Analysis}
\subsection{Extraction of the linewidths}
An FMR mode can be fitted as shown in the main text with the standard model for a hanger-type coupled geometry

\begin{equation}
    S_{21}(\omega)=-1 + \gamma_{\text{ext}}e^{i\theta}/[\gamma_{\text{ext}}+\gamma_{\text{int}}+2i(\omega-\omega_0)].
\end{equation}
In that way, we can extract the internal and external linewidths, $\gamma_{\text{int}}$ and $\gamma_{\text{ext}}$. 
\subsection{Extraction of the coupling}
To extract the effective coupling $g_{\text{eff}}=\sqrt{2}V_k\beta$ from the two-tome measurement datasets as a function of the increasing pump strength, we use a general formula for the power transmission of two strongly coupled modes \cite{zhang2014strongly}:
\begin{equation}
    S_{21}=1+\frac{\gamma_{\text{ext}}}{i(\omega_p-\omega_0)-(\gamma_{\text{ext}}+\gamma_{\text{int}})+\frac{g_{\text{eff}}^2}{i(\omega_p-\omega_d)-\gamma_d}}.
    \label{fittin_eq}
\end{equation}
Here, $\gamma_{\text{ext}}$ and $\gamma_{\text{int}}$ denote the internal and external losses of the resonantly driven $k=0$ mode, such that the total loss contribution introduced in the theory above is $\gamma_0=\gamma_{\text{ext}}+\gamma_{\text{int}}$. The two frequency terms indicate the detuning between the probe at $\omega_p$ and the magnon resonances at $\omega_0$ and $\omega_d$. Since the data are taken at $\Delta=0$, it follows that $\omega_0=\omega_d$.
\begin{figure}[hbt!]
\centering
\includegraphics[width=0.9\linewidth]{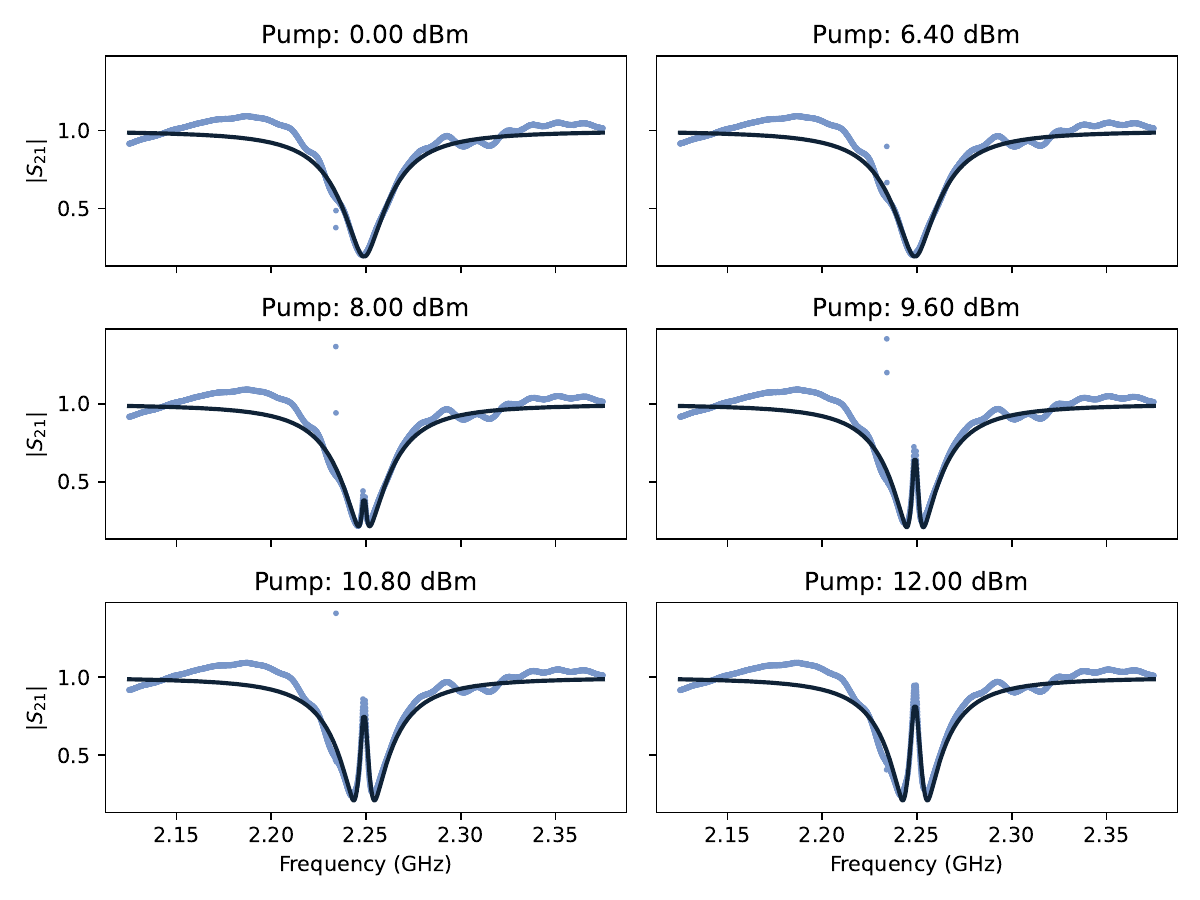}
\caption{Microwave absorption spectra $|S_{21}|$ as a function of frequency for different RF input power applied on resonance. The plots show the emergence of a splitting that increases with power. Blue curves represent the fits of Eq. \ref{fittin_eq}.}
\label{splitting_fits}
\end{figure}
The high density of magnetostatic modes and their respective narrow frequency separation in the measured spectrum make it very difficult to fit the raw data with an ideal Lorentzian response. To account for this, after filtering out the data points corresponding to the strong drive tone and its image leakage on the VNA, we use a polynomial function to fit and remove the background from the measured $S_{21}$. The procedure is similar to the one described in \cite{gely2023apparent}. The least squares fitting method is then used to recover the experimental effective coupling and its corresponding uncertainty. The extracted values of $g_{\text{eff}}$ are then subsequently fitted as a function of power to Eq. \ref{beta_amplitude}. The internal and external losses of the MS modes are derived as explained in the main text, while $\gamma_k$  and the corresponding attenuation of the RF lines in our setup are both kept as free-fitting parameters. Considering standard input-output formalism, we assume that the driving rate $\Omega_d$ is related to the input microwave power as follows:
\begin{equation}
\Omega_d=\sqrt{\frac{P_{\text{in}}\gamma_{\text{ext}}}{2\hbar\omega_0}}.
\end{equation}
Figure \ref{splitting_fits} shows examples of the fitting model to the data as in Figure 2 in the main text. 
\begin{figure}[h!]
    \centering
    \includegraphics[width=\linewidth]{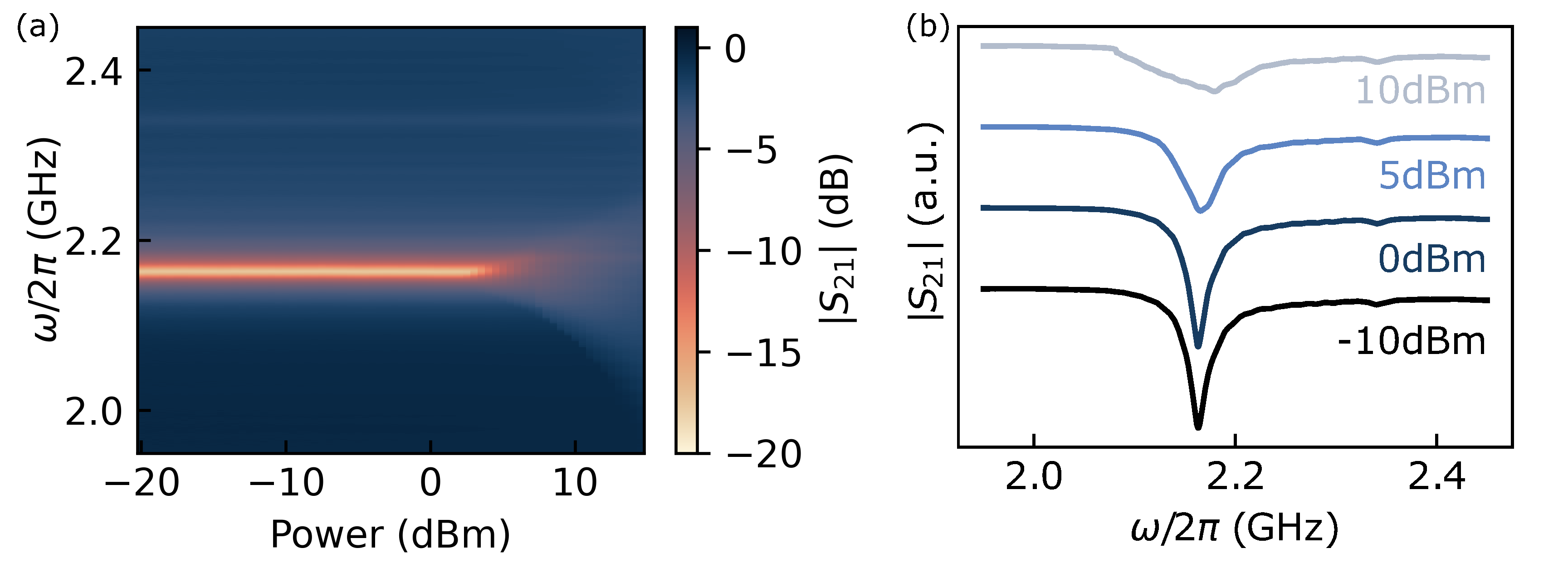}
    \caption{Saturation of the FMR mode and linewidth broadening at high RF power. (a-b) The three-magnon scattering process arising due to Suhl instabilities results in the saturation of the FMR mode above a power threshold.}
    \label{fig:power_broadening}
\end{figure}
\section{FMR Power broadening}
One of the main signatures of the first-order Suhl instability is the saturation of the driven magnon mode with increasing RF power due to the back reaction of the parametrically excited $\pm k$ modes \cite{l2012waveturbulence}. As a consequence, the response of the MS mode will get broader and broader as the power is increased until its line shape completely deviates from a Lorentzian. Fig. \ref{fig:power_broadening} (b) shows how the FMR response broadens as a function of input power, until in the high power limit the absorption spectrum saturates. 
 
\bibliography{SI}